\documentclass[10pt,graphicx,subfigure,axodraw,cleverref]{paper}
\usepackage{bm}
\usepackage{multirow}
\usepackage{amsfonts}
\usepackage{amssymb}
\usepackage{epsfig}
\usepackage{float}
\usepackage{ulem}
\restylefloat{table}

\usepackage[usenames,dvipsnames]{color}
\newcommand{\beqn}{\begin{eqnarray}}
\newcommand{\eeqn}{\end{eqnarray}}
\newcommand{\be}{\begin{equation}}
\newcommand{\ee}{\end{equation}}

\newcommand{\cl}{{\cal L}}

\def\s1{$s_{\alpha}$}
\def\s2{$s_{\gamma}$}
\def\s3{$s_{\delta}$}
\def\c1{$c_{\alpha}$}
\def\c2{$c_{\gamma}$}
\def\c3{$c_{\delta}$}

\def\45{\overline{45}}
\def\5{\overline{5}}
\def\70{\overline{70}}
\def\50{\overline{50}}

\newcommand{\ov}{\overline }

\def\c{\acute{c}}

\def\45{\overline{45}}
\def\5{\overline{5}}
\def\70{\overline{70}}
\def\50{\overline{50}}

\usepackage{amsmath,amsfonts,amssymb}
\usepackage{hyperref}
\usepackage{cleveref}
\crefname{equation}{Eq.}{Eqs.}
\crefname{figure}{Fig.}{Figs.}
\crefname{table}{Table}{Tables}
\crefname{section}{Section}{Sections}

\usepackage{epsfig}
\usepackage{psfrag}
\usepackage{dsfont}

\def\sm{Standard Model~}

\def\GeV{\rm GeV~}

\def\cl{{\cal L}}

\def\kpc{\,\mbox{kpc}}

\newcommand{\beq}{\begin{equation}}
\newcommand{\eeq}{\end{equation}}

\def\G{\tilde G}

\begin{document}

\baselineskip 18pt
\begin{titlepage}

\begin{flushright}
\end{flushright}

\begin{center}
{\bf {\ {\textsf{\Large{
    }}}}}

\vskip 0.5 true cm \vspace{0cm}
\renewcommand{\thefootnote}
{\fnsymbol{footnote}}

{\bf \Large   A Stronger Case for Superunification Post Higgs
Boson Discovery
 }\\~\\
Pran Nath$^a$\footnote{Email: nath@neu.edu} and Raza M. Syed$^{a,b}$\footnote{
Email:  rsyed@aus.edu}
\end{center}
\begin{center}
{\noindent
$^a$\textit{Department of Physics, Northeastern University,
Boston, MA 02115-5000, USA} \\
$^b$\textit{Department of Physics, American University of Sharjah,
P.O. Box 26666,
Sharjah, UAE\footnote{Permanent address}}}\\
\end{center}

{
\vskip 1.0 true cm \centerline{\bf Abstract}
{
Supersymmetry and more specifically supergravity grand unification allow one to extrapolate
physics from the electroweak scale up to the grand unification scale consistent with electroweak data.
Here we give a brief
overview of their current status and show that the case for supersymmetry is stronger
as a result of the  Higgs boson discovery with a mass measurement at $\sim 125$ GeV
consistent with the supergravity grand unification prediction that the Higgs boson mass lie
below 130 GeV.
 Thus the discovery of the Higgs boson
 and the measurement of its mass provide a further  impetus for the search for
 sparticles to continue at the current and future colliders.
  }}

\medskip
\noindent

\end{titlepage}


\tableofcontents


\section{ Introduction \label{sec1}}

The standard model of particle physics~\cite{Glashow:1961tr}
 based on the gauge group $SU(3)_C\otimes SU(2)_L\otimes U(1)_Y$
and three generations of quarks and leptons is a highly successful model at low energies up to the  electroweak scale.
One of the basic elements of the model is that it is anomaly free. Specifically, the quarks and the leptons have
the  $SU(3)_C$, $SU(2)_L$, $U(1)_Y$  quantum numbers  so that
$q(3,2, \frac{1}{6}),\   u^c (\bar 3, 1, -\frac{2}{3}), \  d^c(\bar 3, 1, \frac{1}{3})$, $L (1,2, -\frac{1}{2}), \   e^c (1, 1, 1)$,
where the first two entries refer to the $SU(3)_C, SU(2)_L$ representations, and the last entry refers to the
hypercharge defined so that $Q= T_3+ Y$. The anomaly free condition in this case implies that
one has
$\sum_i f_i Y_i=0$,
where $f_i$ is a product of multiplicity and color factor.  Here one generation of quarks and
leptons exactly  satisfies the  anomaly cancellation condition. The interesting phenomenon  is  that
while the leptons have integral charge, $Q=-1$ for charged lepton, $Q=0$ for the neutrino,
 the quarks have fractional charge,
 $2/3$ for the up quarks and $-1/3$ for the down quark. The charge assignment appears
 intriguing and leads one to ask if there exist a larger framework within which one may understand
 such charge assignments. Such a framework must be more unified and exist at a larger scale.
 There are other aspects which point to the possibility that a more unified framework may
 exist such as the product nature of standard model group. Here it requires three gauge coupling
 constants $g_3, g_2, g_Y$ to describe the interactions. One may speculate if they are remnants
 of a single coupling.  Such issues were the subject of investigations in the early seventies.
 Thus in 1973-74 Pati and Salam ~\cite{Pati:1974yy,Pati:1973uk} 
 proposed that the standard model was remnant of the
 group $G(4,2,2)\equiv SU(4)\otimes SU(2)_L\otimes SU(2)_R$.  Here leptons and quarks are
 unified with the leptons arising as a fourth color.  Soon after the work of ~\cite{Pati:1974yy}
 Georgi and Glashow~\cite{Georgi:1974sy}
  proposed the group $SU(5)$
 which in addition to unifying the quarks and  the leptons, also unifies the gauge coupling constants.
  A group which gives an even greater unification was proposed
  subsequently.  This is the group $SO(10)$~\cite{georgi-so10}.
   It has the benefit
  of unifying a full generation of quarks and leptons in one irreducible representation of $SO(10)$.
    The general criteria for grand unification is that one
    needs those  unification groups  which have chiral representations.
       Here the relevant groups are
       $SU(N); ~SO(4N+2), N\geq 1; ~E_6$.
       As noted
             one of the constraints on model building is that of anomaly cancellation. Here
    the groups $E_6$ and $SO(4N+2), N\geq 1$ are automatically anomaly free while for $SU(N)$ one needs
    combinations of representations which are anomaly free.

    However, as is well known non-supersymmetric models have a serious  fine tuning problem~\cite{Gildener:1976ih}.
    Quantum loop corrections to the Higgs boson mass-squared give contributions which are
    quadratically divergent in the cutoff. In grand unified theories that cutoff would be the GUT
    scale $\sim 10^{16}$ GeV, which is much larger than the electroweak scale. A cancellation of
    the loop term would require a fine tuning of one part in $10^{28}$.
    The cancellation of the quadratic divergence occurs naturally in
    supersymmetry~\cite{Wess:1974tw}.
        It is desirable then
    to formulate unified models using supersymmetry. One persistent problem here
    concerns breaking of supersymmetry which is essential for building a viable phenomenology.
    This is problematic in global supersymmetry.  In order to break supersymmetry in a
    phenomenologically viable way one needs local supersymmetry/supergravity
    \cite{Nath:1975nj,Arnowitt:1975xg,Freedman:1976xh}
    (For a more extensive discussion
    see ~\cite{Nath:2016qzm}).
        Indeed grand unified models based on local supersymmetry provide the appropriate framework
    for unifying the strong and the electroweak interactions~\cite{Chamseddine:1982jx}.
    Gauge coupling unification is an important touchstone of unified model~\cite{Georgi:1974yf}.
    An important success of  supersymmetry models is the unification of gauge coupling constants
 consistent with LEP data ~\cite{g1,g2,g3,g4}.  We note in passing that Planck scale physics could affect
 the predictions at the grand unification scale~\cite{Hill:1983xh}
  (see \cite{g5,g6}).
       Further, a significant feature of supergravity grand unification is that it is also the appropriate vehicle
    for the analysis of
    string based models since supergravity is the low energy limit of strings, i.e., at scales $E<M_{\rm Pl}$
    (see, e.g.,\cite{Green:1987sp}).

    The outline of the rest of the paper is as follows:   In section \ref{sec2} we discuss the first works
    on unification beyond the standard model. These include the quark-lepton unification and the unification of
    gauge coupling constants. As noted the group $SO(10)$ is now the preferred  unification group for the unification of
    the electroweak and the strong interactions. This is discussed in section \ref{sec3}.
    In section \ref{sec4}    we discuss an alternative possibility
    for a unifying group, i.e., $E_6$.
  The flavor puzzle which relates to the hierarchy of quark and lepton  masses and mixings is discussed in
section \ref{sec5}.   In section \ref{sec6} we discuss supergravity grand unification which  provides  the
  modern framework for realistic analyses of grand unified models and allows one to extrapolate physics
  from the electroweak scale to the grand unification scale. Unification in strings is discussed in
section \ref{sec7}. One of the hallmarks of grand unified models is the prediction for the existence of
monopoles and we discuss it in section \ref{sec8}.
We note that  no signature of monopole has thus far been detected and confirmation of its existence
remains an outstanding experimental question.
Since grand unification implies quark-lepton unification, another important prediction of grand unification  is the
decay of the proton. We discuss the current status of proton stability in   GUTs and
strings in section \ref{sec9}.
  In section \ref{sec10} we argue that the discovery of the Higgs boson mass at $\sim 125$ GeV
lends further support for SUSY/SUGRA/Strings, and consequently for the discovery of sparticles at colliders.
The role of future colliders for testing superunification is discussed in section \ref{sec11}. Conclusions are given
in section \ref{sec12}.

\section{$G(4,2,2)$ and $SU(5)$ unification \label{sec2}}
As mentioned in section 1
a significant step toward unification beyond the Standard Model was taken by  Pati and Salam~\cite{Pati:1974yy}
in 1974 when they proposed an extension of the \sm gauge group to the group
 $ G(4,2,2)\equiv SU(4)\otimes SU(2)_L\otimes SU(2)_R$. Here
 $(4,2,1)$ and $(\overline{4}, 1, 2)$  representations of $G(4,2,2)$  contain  a full
generation of quarks and leptons. Since quarks and leptons reside in the same multiplet,
 $G(4,2,2)$  represents unification of quarks and leptons. This phenomenon has a direct consequence
 in that it allows conversion of quarks into leptons and thus one might expect the proton to
become unstable and decay. The feature above is shared by essentially all unified models and thus proton lifetime limits
act  as a strong constraint on unified models of particle unifications. We also note that in  $(4,2,1)+ (\overline{4}, 1, 2)$
 one has
one more particle, i.e., $\nu^c$, which does not appear in the standard model. $\nu^c$  enters in the
so called seesaw mechanism~\cite{seesaw}
that gives mass to the neutrinos.
In  $G(4,2,2)$  the charge operator takes the form
$Q_{em} = T_{3L} + T_{3R} + \frac{B-L}{2}$, where $T_{3L}$ and $T_{3R}$ are the generators of $SU(2)_L$ and
$SU(2)_R$ and $B$ and $L$ are baryon and lepton numbers.
To break the $G(4,2,2)$ symmetry one introduces heavy Higgs representations $(4,2,1)_H \oplus (\bar 4, 1,  2)_H$.
 In this case when $\nu_H$ and $\nu^c_H$ develop VEVs, i..e,
$<\nu_H>=<\nu^c_H>= M_{\nu_H}$,
 $G(4,2,2)$ breaks to the \sm gauge group  and the charge operator takes the familiar form
$Q_{em} = T_{3L} + Y$ where
$Y= T_{3R} + \frac{B-L}{2}$.  The group $G(4,2,2)$ can be broken further down to  $SU(3)_C\otimes U(1)_{em}$
by use of $(1,2,2)$ Higgs representation. A comprehensive review of $G(4,2,2)$ was recently given in
~\cite{Pati:2017ysg}.

$SU(5)$\cite{Georgi:1974sy} is the lowest rank unified group which can accommodate the standard model gauge group.
Here a full generation of quarks and leptons can be accommodated in its  representations
$\bar 5 \oplus 10$.  Under the \sm gauge group they decompose so that $5=(3, 1, -\frac{1}{3}) \oplus (1,2, \frac{1}{2})$
and
$10=(3,2, \frac{1}{6}) \oplus (\bar 3, 1, -\frac{2}{3}) \oplus (1,1,1)$,
where one identifies $(\nu_L, e_L)$ with $(1,2, -\frac{1}{2})$,  $e^c_L$ with  $(1,1,1)$,  $(u_L, d_L)$ with  $(3,2, \frac{1}{6})$,   $u^c_L$ with $(\bar 3,1,-\frac{2}{3})$ and  $d^c_L$ with
  $(\bar 3,1,\frac{1}{3})$. The GUT symmetry is broken by a 24-plet of heavy field $\Sigma^i_{j}~ (i,j=1,2,\cdots, 5)$
  which breaks $SU(5)$ down to the \sm gauge group. To break the symmetry further to the residual gauge
  group $SU(3)_C\otimes U(1)_{em}$  one introduces a $5$ plet of Higgs $H$ in the non-supersymmetric case.
  Here one immediate issue concerns the so-called doublet-triplet problems, i.e., how to keep the
  doublet of the 5-plet of Higgs light while making the triplet of the 5-plet superheavy.
A concrete way to see this problem is to consider the scalar potential
$V= M^2 {\rm Tr} \Sigma^2 +  \frac{\lambda_1}{2} {\rm Tr}(\Sigma^4) +
 \frac{\lambda_2}{2} ({\rm Tr} \Sigma^2)^2+ \mu {\rm Tr} (\Sigma^3)
+  \frac{1}{2}\lambda_3 H^{\dagger} H {\rm Tr}(\Sigma^2)
+ \frac{1}{2} \lambda_4 H^{\dagger} \Sigma^2 H + \frac{\lambda}{4} (H^{\dagger}H)^2$.
 In order to break $SU(5)$ down to the \sm gauge group we need to have  the VEV formation of $\Sigma$
so that  $<\Sigma> ={\rm  diag} ( 2, 2, 2, -3, -3) v$.
Here the spontaneous breaking of the symmetry gives the constraint
$M^2+ (7 \lambda_1 + 30 \lambda_2) v^2 - \frac{3}{2} \mu v =0$.
The breaking generates a mass for the Higgs doublet which is superheavy whereas electroweak symmetry
breaking requires that the Higgs doublet be light. In order to achieve a light Higgs doublet we need
 the constraint  $10 \lambda_3+ 3 \lambda_4=0$. This
constraint must be satisfied to  one part in $10^{28}$, which is a high degree of fine tuning.

  In $SU(5)$ GUT the hypercharge coupling $g_Y$  is related to the $SU(5)$ invariant coupling $g_5$ so that
$g_Y = \sqrt{{3}/{5}} g_5$. Thus $SU(5)$ predicts the weak angle
 at the GUT scale so that
$\sin^2\theta_W = {g^2_Y}/({g_2^2 + g_Y^2}) ={3}/{8}$, where we set  $g_2=g_5$.
 One of the problems of the minimal $SU(5)$  model is that it
generates undesirable relationships among the quark and the lepton masses.
Thus  consider the $SU(5)$ Yukawa couplings
${\cal L}_Y=  h_d  \bar \psi^c_{i} \psi^{ij} H^{\dagger}_j
+  h_u \epsilon_{ijklm} \bar \psi^{c ij} \psi^{kl} H^m + h.c.$,
where  $ \psi^c_i$ and $\psi^{ij}$ are the $\bar 5$-plet and  10 -plet of fermions. In the above we have
suppressed the generation index. More explicitly each generation will have their own Yukawa couplings
$h_d$ and $h_u$. After spontaneous breaking when  the Higgs doublet gets a VEV one finds the following mass
relation
$m_e = m_d, ~m_{\mu} = m_s, ~m_b = m_{\tau}$.
For the first two generations one has at the GUT scale the equality
${m_e}/{m_{\mu}} = {m_d}/{m_s}$. These ratios are independent of the scale to one loop order
and they hold at the electroweak scale to this order.  However, the relation is badly in conflict with data.
For the third generation $m_b=m_\tau$ also does not quite work for the non-supersymmetric case
although  it does for the supersymmetric case.
This means that the minimal  $SU(5)$ model is not sufficient to
explain the flavor structure of  the Standard Model.
In $SU(5)$ there are 24 gauge bosons of which 12 are the gauge bosons of the Standard Model consisting of the
gluon, $W^{\pm}, Z, \gamma$. The remaining  gauge bosons  are superheavy leptons-quarks $(X^{4/3},Y^{1/3}),
(\overline{X}^{-4/3}, \overline{Y}^{-1/3})$ where the super-scripts give the charge assignments.
However, non-supersymmetric $SU(5)$ does not produce unification of gauge coupling constants
consistent with LEP data.

 We turn now to the supersymmetric version of $SU(5)$~\cite{Dimopoulos:1981zb}.
Here the superpotential for the minimal supersymmetric  $SU(5)$ is given by
$W = \lambda_1 [\frac{1}{3} {\rm Tr} \Sigma^3 + \frac{1}{2} M {\rm Tr} \Sigma^2]
+ \lambda_2 {H}_{1i} [\Sigma_j^i $+ $2 M'
\delta_j^i]H_2^j
+f_1  H_1\cdot\bar 5\cdot10  + f_2 H_2\cdot10\cdot10$.
The GUT symmetry breaks  as in the non supersymmetric case with $\Sigma$ developing a VEV.
Here as in the non-supersymmetric case a fine tuning is needed to get a light doublet.
Specifically under the constraint $M'=M$ the Higgs doublets are light while the Higgs triplets
are superheavy.
 While a fine tuning is needed in the minimal
$SU(5)$ model to get  light Higgs doublets, it is possible to achieve the doublet-triplet splitting without fine tuning in
non-minimal or extended $SU(5)$ models.
One example is the
so called missing partner mechanism~\cite{Grinstein:1982um,Masiero:1982fe}
 and the other flipped $SU(5)\otimes U(1)$ model~\cite{flipped}.
In the missing partner mechanism one uses a 75-plet representation to break the
GUT symmetry and additionally  $50 +\overline{50}$ representations to give masses to the color triplets.
 The superpotential involving the $50$, $\overline{50}$ and $75$ plet has the form
$W_G = \lambda_1 \overline{ \Phi}_{ijk}^{lm} \Sigma_{lm}^{ij} H_2^k $+$ \lambda_2
{\Phi}_{lm}^{ijk} \Sigma_{ij}^{lm} \overline{H}_{1k}
+ M \overline{ \Phi}_{ijk}^{lm} {\Phi}_{lm}^{ijk} + W_{\Sigma} (\Sigma)$,
where $50 =\Phi^{ijk}_{lm}$,
$\overline{50} =  \overline{\Phi}^{lm}_{ijk}$,
and $75  =  \Sigma_{lm}^{ij}$.
The superpotential  $W_{\Sigma} (\Sigma)$ is  chosen to produce a GUT symmetry breaking when
 $\Sigma_{kl}^{ij}$ develops a non-vanishing VEV of $O (M)$.
The  $SU(3) \otimes SU(2)\otimes U(1)$ decomposition of the  $75$ plet  contains the singlet field $(1,1)(0)$  and its
VEV formation breaks the $SU(5)$ GUT symmetry. The Higgs fields $5$ and $50$-plet have the
$SU(3) \otimes SU(2)\otimes U(1)$ decomposition so that
   $5=(1,2)(3) \oplus (3,1)(-2)$  and  $50 = (1,1)(-12)  \oplus  (3,1)(-2) \oplus (\bar 3,2)(-7)$
$\oplus (\bar 6,3)(-2) \oplus ({6},1)(8)\oplus (8,2)(3)$.
Here it is interesting to note that the $5$-plet contains a Higgs triplet  and similarly $\bar 5$ contains an
anti-triplet, while $50$-plet contains a Higgs triplet and similarly $\overline{50}$ contains a Higgs anti-triplet.
Thus the interaction ${5\cdot50\cdot75}$ gives superheavy masses to the triplets and anti-triplet by
matching the triplet in $5$ to the anti-triplet in $\overline{50}$ and similar tying the anti-triple in $\bar 5$ with the
triplet in $\overline{50}$. On the other hand, $50$ and $\overline{50}$ do not have Higgs doublets and thus
the Higgs doublets of $5$ and $\bar 5$ remain light.

Another method of producing light Higgs doublets while keeping the Higgs triplets heavy is to use the
flipped $SU(5)\otimes U(1)$. As in the case of $SU(5)$ one uses $\bar 5\oplus 10$ plet representations of $SU(5)$
for each generation of quarks and leptons. However, the $u$ and the $d$ quarks as well as $e$ and $\nu$ leptons are
interchanged,  the right handed neutrino $\nu^c$  replaces $e^c$ in the 10-plet representation and $e^c$
appears in  the singlet representation.
To break the GUT symmetry one uses in the Higgs sector $10\oplus\overline{10}$  rather than the 24-plet.
For the breaking of the electroweak symmetry one introduces a  $\bar 5\oplus 5$, i.e., $H_{1}$ and
$H_2$  as for the  standard  $SU(5)$ case. The superpotential for the Higgs sector is then
$W_{\rm flipped} = {W_0(10)} + \lambda_1 \epsilon_{ijklm} H^{ij} H^{kl} H_2^m
+\lambda_2 \epsilon^{ijklm} \bar H_{ij} \bar H_{kl} \bar H_{1m}$ where $W_0(10)$ is the superpotential
that depends only on the $10$ and $\overline{10}$.
 The $SU(3) \otimes SU(2)\otimes U(1)$ branchings of the $10$-plet of $SU(5)$ are given by
 $10= (\overline{3},1) (-4)\oplus (3,2)(1) \oplus (1,1) (6)$.
 When
  the singlet field $(1,1)(6)$ develops a VEV,  the
color triplet in the $5$-plet combines with the color anti-triplet in $10$-plet to become supermassive,
while the $SU(2)$ doublet and color singlet  $(1,2)$ in the $5$-plet has no partner in the
$10$-plet. Thus one gets a natural missing partner mechanism in this case.
The $X,Y$ gauge fields of $SU(5)$  have the $SU(3)_C\otimes SU(2)_L\otimes U(1)_Y$ quantum numbers
$(X,Y)= (3,2,5/6)$, and the charges for them are
$Q_X=4/3, Q_Y= 1/3$.  In contrast the $(X',Y')$ gauge bosons of the flipped $SU(5)\otimes U(1)$ have the quantum numbers
$(X',Y')=(3,2, -1/6)$ so that the charges for $X'$ and $Y'$ are given by $Q_{X'}= 1/3, Q_{Y'} = -2/3$.
 The unusual charge assignment in this case requires that the hypercharge
be a linear combination of $U(1)$ and of the generators in $SU(5)$.  A
drawback of the model is  that it is not fully unified  since
the underlying  structure is a product group.
\section{SO(10) unification \label{sec3}}

 The group $SO(10)$  as the framework for grand unification  appears preferred over $SU(5)$.
 The group $SO(10)$ contains both $G(4,2,2)$ and $SU(5)\otimes U(1)$ as subgroups, i.e.,
 $SO(10)$ has the branchings $SO(10)\to SU(4)_C\otimes SU(2)_L\otimes SU(2)_R$ and
 $SO(10)\to SU(5)\otimes U(1)$.
It possesses a spinor representation  which is $2^5=32$
 dimensional and which splits into $16\oplus \overline{16}$.
 A full generation of quarks and leptons can be accommodated in a single $16$ plet
 representation.  Thus the $16$-plet has the decomposition in $SU(5)\otimes U(1)$  so that
 $16 = 10(-1)\oplus\overline{5}(3)\oplus1(-5)$.
As  noted  the combination $\overline{5} \oplus10$ in $SU(5)$  is anomaly free and further $1(-5)$
 in the $16$-plet decomposition is a right handed neutrino which is a singlet of the standard model gauge group and thus the 16-plet of
matter in $SO(10)$ is  anomaly free.
The absence of anomaly in this case is the consequence of a more general result for $SO(N)$ gauge theories.
Thus in general
anomalies arise due to the non-vanishing of the trace over the product of three
group generators in some given group representation
${\rm Tr} \left(\{T_a, T_b\}T_c\right)$.
For $SO(10)$ one will have
${\rm Tr} \left(\{\Sigma_{\mu\nu}, \Sigma_{\alpha\beta}\}\Sigma_{\lambda\rho}\right)$.
However, there is no invariant tensor to which the above quantity can be proportional  which
then automatically guarantees vanishing of the anomaly for $SO(10)$. This analysis extends to other
$SO(N)$ groups. One exception is $SO(6)$ where there does exist a six index invariant tensor
$\epsilon_{\mu\nu\alpha\beta\lambda\rho}$ and so in this case vanishing of the anomaly is not
automatic.

The group $SO(10)$ is rank 5 where as the standard model gauge group is rank 4. The rank of the
group can be reduced by either using $16\oplus\overline{16}$ of Higgs fields or $126\oplus\overline{126}$ of Higgs.
Since under $SU(5)\otimes U(1)$ one has
$16\supset 1(-5)$
 we see that a VEV formation for the singlet  will reduce the rank of the group. Similarly
$126\supset 1(-10)$ under the above decomposition. Thus when the singlets in
 $16\oplus\overline{16}$ of Higgs or $126\oplus\overline{126}$ get VEVs, the $SO(10)$
  gauge symmetry will break
 reducing its rank.
  However, we still need to reduce the remaining group symmetry to the \sm gauge group.
 For this we need to have additional Higgs fields  such as $45, 54, 210$. Further to get the residual
 gauge group $SU(3)_C\otimes U(1)_{em}$ we need to have $10$ -plet of Higgs fields.
Thus the breaking of $SO(10)$ down to $SU(3)_C\otimes U(1)_{em}$ requires at least three
sets of Higgs representations: one to reduce the rank, the second to break the rest of the gauge
group to the \sm gauge group and then at least one 10-plet to break the electroweak symmetry.
 As discussed above one can do this by a combination of  fields from the set:
  $10, 16\oplus \overline{16}, 45, 54, 120, 126 \oplus \overline{126}, 210$.
 To generate quark and lepton masses we need to couple two $16$-plets of matter
with Higgs fields. To see which Higgs fields couple we expand the product $16\otimes 16$ as a
sum over the irreducible representations of $SO(10)$. Here we have
 $16 \otimes 16 = 10_s\oplus 120_a \oplus {126}_s$,
where the $s(a)$ refer to symmetric (anti-symmetric) under the interchange of the two 16-plets.
 The array of Higgs bosons available lead to a large number  of possible $SO(10)$ models.
 For some of the works utilizing large representations see \cite{so10-largereps}.

 As discussed above the conventional models   have the drawback that one needs
 several sets of Higgs fields to accomplish complete breaking. One recent suggestion to overcome
 this drawback is to use $144\oplus\overline{144}$ of Higgs. This combination can allow one to break
 $SO(10)$ symmetry all the way to $SU(3)_C\otimes U(1)$. This can be seen as follows:
In  $SU(5)\otimes U(1)$ decomposition one  finds that  $144 \supset 24(-5)$ which
 implies that spontaneous symmetry breaking which gives VEV to the \sm singlet of $24$
   would also reduce the rank of the group. Thus in one step one can break $SO(10)$ gauge symmetry
   down to the \sm gauge group. Further, there exist several Higgs doublets in $144$
   which have the quantum numbers of the \sm Higgs and  one may use
   fine tuning to make one of the Higgs doublets light which is needed to give masses to the quarks and the leptons.
    A VEV formation that breaks the $SO(10)$ symmetry can be achieved by taking a combination of dimension
    two and dimension four terms in the potential so that
 $V=-M^2 {\rm Tr}(\Sigma \Sigma^{\dagger})
+\frac{\kappa_1}{2} {\rm Tr}(\Sigma^2 \Sigma^{\dagger 2})
 +\frac{\kappa_2}{2} ({\rm Tr}(\Sigma \Sigma^{\dagger}))^2$
 $+\frac{\kappa_3}{2} {\rm Tr}(\Sigma^2) {\rm Tr}(\Sigma^{\dagger 2})
+\frac{\kappa_4}{2} {\rm Tr}(\Sigma \Sigma^{\dagger}  \Sigma \Sigma^{\dagger}   )$.
Symmetry breaking allows for local minima which lead to the \sm gauge group for the vacuum structure so that
 $<\Sigma >= <\Sigma^{\dagger}> =v ~diag(2,2,2, -3, -3)$, where
 $v^2 = { M^2}/({ 7(\kappa_1+\kappa_4)+ 30 (\kappa_2 + \kappa_3)})$.
 This local minimum will be a global minimum for some  some range of the parameters of the potential
which implies that  spontaneous symmetry breaking does indeed break $SO(10)$ down
 to $SU(3)_C\otimes SU(2)_L\otimes U(1)_Y$.
Identical conclusions can be arrived at for the case of
supersymmetric $SO(10)$ model.  Here the potential
 will become the superpotential (with
$\Sigma^\dagger$ replaced by a chiral superfield $\overline{\Sigma}$), the
couplings $\kappa_i$ will have
inverse dimensions of mass, and the mass term $M^2$ will be
replaced by $M$. The analysis of~\cite{Babu:2005gx,Babu:2006rp} shows that in SUSY $SO(10)$ a
$144\oplus\overline{144}$ pair of chiral superfields do indeed break $SO(10)$
in one step down to the supersymmetric standard model gauge group.  Techniques for the computation of
$SO(10)$ couplings using oscillator algebra~\cite{Mohapatra:1979nn}
 and $SU(5)\otimes U(1)$ decomposition  are discussed in
 ~\cite{Nath:2001uw,Nath:2001yj,Nath:2003rc,Nath:2005bx}. Specifically this technique is very useful
for analyses involving vector-spinors, i.e., $144\oplus\overline{144}$. For alternative techniques see
 ~\cite{Aulakh:2002zr}.

 As mentioned in sec \ref{sec2}  grand unified models typically have the doublet-triplet problem.
 In $SU(5)$ aside from fine tuning two other avenues to overcome this problem were mentioned,
 one being the missing partner mechanism and the other flipped $SU(5)\otimes U(1)$. In $SO(10)$
 one early suggestion to resolve the doublet-triplet problem is the missing VEV method
  where using a $45$-plet of Higgs fields, one breaks the
 $SO(10)$ gauge symmetry  in the $(B-L)$--preserving
direction, which results in generation of Higgs triplet  masses  while the Higgs doublets
from a $10$-plet remain massless.
Another possibility is a missing partner mechanism in $SO(10)$ which in spirit is similar to the one for $SU(5)$.
In  \cite{Babu:2006nf} a  missing partner mechanism
 for $SO(10)$ was constructed and the heavy fields used were $126\oplus\ov{126}\oplus210$ along
 with a set of light fields.  The reason for this choice is to simulate the missing partner mechanism
 of $SU(5)$ in the following way:  $126\oplus\ov{126}$ are chosen because they
  contain  $50\oplus\ov{50}$ of $SU(5)$ and $210$ is chosen  because it contains the $75$ of $SU(5)$.
  This parallels the analysis in $SU(5)$ and leads to a pair of light Higgs doublets and  heavy Higgs triplets.

 In~\cite{Babu:2011tw}  a more comprehensive analysis of the missing partner mechanism for SO(10) was given.
 Here in addition to the missing partner model consisting of
(i) Heavy  $\{\mbox{{$126\oplus\ov{126} \oplus  210$}}\} $
+ Light  $\{\mbox{{$2 \times 10 \oplus  120$}}\} $ one has others:
(ii)  Heavy  \{\mbox{{$126\oplus\ov{126} \oplus 45$}}\} + Light  \{\mbox{{$ 10 \oplus  120$}}\},
(iii) Heavy  \{\mbox{{$126\oplus\ov{126} $}}\} + Light  \{\mbox{{$ 10 \oplus  120$}}\},
(iv) Heavy  \{\mbox{{$560\oplus\ov{560} $}}\} + Light  \{\mbox{{$  2 \times 10 \oplus  320$}}\}.
Models (i), (ii) and (iii) are anchored in the heavy fields $126+\ov{126}$ since they contain
the $50\oplus\ov{50}$ of $SU(5)$.
However,  model (iv)  is anchored by a pair of $560\oplus\ov{560}$ Higgs fields which also
contain $50\oplus\ov{50}$ of $SU(5)$.  Interestingly these are the next higher dimensional representations
in $SO(10)$ after $126\oplus\ov{126}$ which contain an  excess of color triplets over
$SU(2)_L$ doublets. Further, it turns out that $560$ also contains in the $SU(5)$ decomposition
the $SU(5)$ representations  $1(-5)\oplus24(-5)\oplus75(-5)$. After spontaneous breaking these fields
acquire VEVs. They reduce the rank of the group from 5 to 4 and further break the gauge
symmetry down to the symmetry of the  \sm gauge group.  That means that the $SO(10)$
gauge symmetry breaks to the \sm gauge group in one step. This is reminiscent of
the unified Higgs case discussed earlier in this section where  $144\oplus\overline{144}$
break the $SO(10)$ gauge symmetry in one step.

 Next we explain how the missing partner
mechanism works in this case. To this end we exhibit the complete $SU(5)\otimes U(1)$ decomposition of
$560$. Here one has
$560=1(-5)\oplus\overline{5}(3)\oplus\overline{10}(-9)\oplus10(-1)_1\oplus10(-1)_2\oplus24(-5)+40(-1)$
$+45(7)\oplus\overline{45}(3)\oplus\overline{50}(3)\oplus\overline{70}(3)\oplus75(-5)\oplus175(-1)$.
Regarding the light sector it turns out that we have a unique choice in this case, i.e.,
the light fields must be in  $(2\times 10\oplus320)$ representations.  A very stringent condition for the
missing partner mechanism to work is that all the exotic fields must become heavy and thus
in the entire array of Higgs fields only a pair of Higgs doublet fields must survive and remain light.
To exhibit that this indeed is the case let us consider the $SU(5)\otimes U(1)$ decomposition
of $320$ and of $10$'s. For the $320$ we have
$320= 5(2) + \overline{5}(-2) \oplus 40(-6) \oplus \ov{40}(6) \oplus 45(2)\oplus\ov{45}(-2) \oplus 70(2) \oplus\ov{70}(-2)$,
and for the $10$-plet we have $10= 5(2) \oplus \overline{5}(-2)$.
We also note that
$45$ of $SU(5)$ under $SU(3)_C \otimes SU(2)_L \times U(1)$  has the decomposition so that
$45 = (1,2)(3)\oplus(3,1)(-2)\oplus(3,3)(-2)\oplus(\ov{3},1)(8)+
(\ov{3},2)(-7)\oplus(\ov{6},1)(-2)\oplus(8,2)(3)$, which exhibits  the  doublet/triplet content of  the 45-plet.
One now finds that the exotic light modes become superheavy by mixing with the heavy exotics
in $560\oplus\overline{560}$ and only a pair of light Higgs doublets remain. Also remarkably the gauge group
$SO(10)$ breaks directly to the \sm gauge group right at the GUT scale so that we have one step breaking
of the gauge symmetry.

   Higher dimensional operators appear in effective theories and
   allow one to explore the nature of physics beyond the standard model. They have been explored in significant depth in the literature.
 These operators also include the ones that violate $B-L$. Such operators appear in $n-\bar n$ oscillations and more
recently they have gained further attention as they may be helpful in generating
baryogenesis~\cite{Enomoto:2011py}.
It is interesting to investigate the type of $B-L$ violating operators  that arise in grand unified theories.
The minimal $SU(5)$ grand unification under the assumption of $R$ parity conservation and renormalizable
interactions does not possess any $B-L$ violating operators. However, $SO(10)$ models
do generate $B-L$ violation. Recently an  analysis of $B-L=-2$ operators has been given in~\cite{Nath:2015kaa}
for a class of $SO(10)$ models where there is doublet-triplet splitting using the missing partner mechanism~\cite{Babu:2011tw}
(for related work see ~\cite{Aulakh:2017aeo}).  The $\Delta (B-L)=\pm 2$ operators lead to some unconventional proton decay modes
such as $p\to \nu \pi^+, n\to e^- K^+$ and $p\to e^- \pi^+$.   The $\Delta (B-L)=\pm 2$ operators  also allow for GUT scale
baryogenesis. The baryogengesis produced by such operators is not wiped out by sphaleron processes and
survives at low temperatures~\cite{Enomoto:2011py}.

   \section{$E_6$ unification \label{sec4}}

   Among the exceptional groups only $E_6, E_7, E_8$ are possible candidates for unification.
   However, $E_7, E_8$ are eliminated as they do not have chiral representations which
   leaves $E_6$ as the only possible candidate for unification among the exceptional groups.
 The lowest representation of  $E_6$
 is $27$-plet and $27\otimes \overline{27}=  1\oplus78\oplus650$ where $78$ -plet is the adjoint representation.
 One notes that
$27\otimes 27 =  \overline{27}_s \oplus \overline{351}_a \oplus \overline{351'}_s $. This result leads to the remarkable
feature of $E_6$ models that the only cubic interaction one can have for 27 is
$W_{27}= \lambda \ 27\otimes 27\otimes 27$.
 $E_6$ has many possible breaking schemes such as
 $E_6\to SO(10) \otimes U(1)_\psi, ~SU(3)_C\otimes SU(3)_L \otimes SU(3)_R, ~SU(6)\otimes SU(2)$.
 One of the most investigated symmetry breaking schemes is $SU(3)^3$. One can exhibit the spectrum of
 $E_6$ in terms of representations of $SU(3)^3$  so that
 $27 = (1,3,\bar 3) \oplus (3,\bar 3, 1) \oplus (\bar 3, 1,3)$
 and thus the  particle content of the model consists of nonets of leptons, quarks and conjugate
 quarks where $L=(1,3,\overline{3})$ ,  $Q=(3,\overline{3}, 1)$ and $Q^c=(\overline{3}, 1,3)$.
The symmetry of the group can be broken by the combination of Higgs fields
${27}\oplus{\overline{27}}\oplus{351'}\oplus{\overline{351'}}$.
 Extensive analyses exist in the literature and  it is shown that with appropriate symmetry breaking schemes
 $E_6$ can produce a low energy theory
 consistent with data (see, e.g., ~\cite{Babu:2015psa,Bajc:2013qra}).
Investigation of $E_6$ as the unification group within string theory  has a long history.  In models of this type $E_6$ is broken down to the
standard model gauge group by a combination of flux breaking and breaking by VEVs of Higgs fields.
In one breaking sequence one has
$E_6\rightarrow SO(10)\otimes U(1)_{\psi}$, ~$SO(10)\rightarrow SU(5)\otimes U(1)_{\xi}$,
~$SO(10)\rightarrow SU(3)_C\otimes SU(2)_L\otimes U(1)_Y$.  Phenomenology of such breaking and of other scenarios leads to
some distinctive signatures.  More recently $E_6$ unification has also been investigated
within F-theory (see e.g., \cite{Callaghan:2013kaa} and the references therein).

\section {The Flavor  Puzzle\label{sec5}}
 The observed structure of quarks and leptons  exhibits some very interesting  properties.
 There are at least two broad features which may be summarized as follows:
 (i) The quarks and leptons show a  hierarchy in masses for different flavors; (ii)
 The mixing among quarks in going from flavor to mass diagonal basis  is small, while
 the mixing among neutrinos is large.
 The
   constraints on the neutrino mixing come from the solar and the atmospheric neutrino oscillation
  data which is sensitive only to the differences of neutrino mass squares, i.e., $\Delta m_{ij}^2= |m_i^2-m_j^2|$.
 A fit to the  neutrino data gives~\cite{nudata}
$\Delta m^2_{sol} =(5.4-9.5)\times 10^{-5}~ eV^2$,
$\Delta m^2_{atm} =(1.4-3.7)\times 10^{-3}~ eV^2$,
$\sin^2\theta_{12}=(0.23-0.39)$,  $\sin^2\theta_{23}=(0.31-0.72)$,
$\sin^2\theta_{13}<0.054$, where
$\Delta m^2_{sol}=||m_2|^2-|m_1|^2|$,  $\Delta m^2_{atm}=||m_3|^2-|m_2|^2|$.
Here
 the mixing angles $\theta_{12}$ and $\theta_{23}$ are  large
while $\theta_{13}$ is small. The constraints on the absolute value of the masses themselves arise from neutrino less double beta decay and from cosmology. Thus from cosmology one has
$\sum_i|m_{\nu_i}|< (0.7-1) eV$.

 One avenue to decode the flavor structure is to assume that  while it  looks complex at low scales, there could  be
 simplicity at high scales.
  An example of this is the ratio of the mass of the $b$ quark vs the mass of the $\tau$ lepton which
  is approximately three at low energy but one can explain this ratio starting from the equality of the $b$ and $\tau$
  Yukawa couplings  at the GUT scale. This result  holds in supergravity GUTs but not in non-supersymmetric
  unification.  In fact, one can also explain the ratio of the top mass to the
 b-quark mass starting with equality of the bottom and the top Yukawa coupling at the GUT scale if one
 assumes large $\tan\beta$~\cite{Ananthanarayan:1991xp}.
 One of the early works in correlating low energy data on quark and lepton
 flavors with textures at the GUT scale is by Georgi and Jarlskog (GJ)~\cite{Georgi:1979df}.
  Aside from the GJ textures there are several other textures  that can
 generate the desired flavor structure at low energy (for early works see \cite{Froggatt:1978nt,harvey}).
  The question then is what are the
 underlying models which can produce the desired textures.
 One possibility is that they  arise from cubic and higher dimensional operators.
 In the analysis of ~\cite{Nath:1996ft}
 it was shown that in $SU(5)$,
 higher dimensions operators in  an expansion in $\Sigma/M$, where  $\Sigma$ is the $24$-dimensional field
 that breaks the GUT symmetry, can produce the GJ textures.
 This analysis also revealed that similar textures exist in dimension five operators
 which enter in proton decay  which are
 different from the ones that appear in the Yukawa couplings.

Within renormalizable interactions one approach is to expand the superpotential in all allowed
couplings and try to fit the data with the assumed set of couplings. For example, for $SO(10)$
one could use the Higgs fields $10_s, 120_a, \overline{126_s}$ and assume general  form of
Yukawas consistent with these couplings, use renormalization group  evolution from the GUT scale down to
low energy and fit the quark and lepton masses and mixings~ \cite{Dutta:2009ij}.
Other approaches involve flavor symmetries. One of the common flavor symmetry used
is $S_4$ of which there are a large number of variants., see. e.g., \cite{Dutta:2009bj} and
\cite{Bjorkeroth:2017ybg}  and the references therein.

 The flavor structure can be understood in another way in a class of   $SO(10)$ models
 within the  unified Higgs sector~\cite{Babu:2005gx,Nath:2005bx}.  As discussed  in
 sec \ref{sec3} in
 unified Higgs models, one uses a  single pair of
vector--spinor representation $144+\overline{144}$  which breaks the $SO(10)$ gauge symmetry
to the standard model gauge group.  Here the matter fields can couple to the Higgs sector only at the quartic level,
i.e., the interaction involves terms such as $16\cdot 16\cdot 144\cdot 144$. This must be suppressed by a heavy mass.
For this reason the Yukawa couplings arising from this interaction are naturally small and  can provide an appropriate
description of the masses and mixings of the first two generations.
However, for the third generation, one needs cubic couplings
and one can obtain much larger couplings in a natural way by including additional  matter in $10$, $45$ and $120$.
The additional matter representations
 allow one to have couplings of the type $16\cdot144\cdot10$, $16\cdot144\cdot 45$,
and $16\cdot144\cdot120$. Specifically using $16\cdot144\cdot 45$, and $16\cdot144\cdot120$
one finds that  cubic couplings of size appropriate for the third generation arise~\cite{Nath:2005bx}.
One can obtain $b-\tau$ as well as $b-t$ unification even at low values of $\tan\beta$. The formalism  also  correctly
generates the charged lepton and neutrino masses which arise from a type I see-saw mechanism\cite{Nath:2005bx}.
There are also a variety of other approaches. For some recent ones see~\cite{Deppisch:2016jzl}.

\section{Supergravity grand unification \label{sec6}}

{
As mentioned in the introduction,  supersymmetry which is a global symmetry, cannot be broken
in a phenomenologically viable fashion.  Supergravity grand unification overcomes
this problem and allows one to build realistic models with spontaneous breaking of supersymmetry
which lead to sparticles with calculable masses which can be searched for at colliders.
 The formulation of  supergravity grand unification utilizes the framework of applied suerrgravity
  where one couples an arbitrary number of $N=1$ chiral superfields, which we denote by $Z$, to $N=1$ vector superfield
 belonging to the adjoint  representation of the  gauge group and then couple them to
 $N=1$ supergravity~\cite{Chamseddine:1982jx,Cremmer:1978iv,Cremmer:1982en, Nath:1983fp}.

The applied $N=1$ supergravity lagrangian  depends
 on three arbitrary functions consisting of the superpotential $W(Z)$,   the Kahler potential potential
  $K(Z, Z^\dagger)$ and the gauge kinetic energy function $f_{\alpha\beta}$ which
   transforms like the symmetric product of adjoint representations.
However,
before discussing the breaking of supersymmetry
we begin by considering the breaking of a grand unified group with no breaking of supersymmetry
in the framework of supergravity grand unification.  As an illustration let us consider the simplest case
where one has a $24$-plet of $SU(5)$ field $\Sigma_a^{~b}$ ($a,b=1-5$)
and a superpotential that is given by
  $W_{\sigma}= [
   \frac{1}{2} M Tr \Sigma^2 + \frac{1}{3} Tr \Sigma^3 ]$.  After a
   spontaneous breaking of $SU(5)$ occurs and $\Sigma_a^{~b}$ develops a VEV,   one
  has three possibilities:
   (i)  $<\Sigma_{~b}^{a}>= 0$,
~$(ii)  <\Sigma_{~b}^{a}>= \frac{1}{3} M\ \left[ \delta_{~b}^{a} - 5 \delta_{~5}^{a} \delta_{~b}^{5}\right]$,
  and $(iii) <\Sigma_{~b}^{a}>=  M\ \left[2 \delta_{~b}^{a} - 5 (\delta_{~5}^{a} \delta_{~b}^{5}+
  \delta_{~4}^{a} \delta_{~b}^{4})\right]$.
Here (i) gives no breaking of gauge symmetry, (ii) gives the breaking of $SU(5)$ to $SU(4) \otimes U(1)$, and
 (iii) gives the breaking to  $SU(3)\otimes SU(2)\otimes U(1)$.

 In global supersymmetry these are flat directions
and the potential vanishes for all the three cases. Thus all three vacua are degenerate.
However, for the case of supergravity the potential does not vanish and one finds that the potential at the minimum
is given by ~\cite{Chamseddine:1982jx}
 $V_{0}(\Sigma_0, \Sigma_0^*) = -\frac{3}{4} \kappa^2  e^{(\frac{1}{2}\kappa^2 \Sigma_0 \Sigma_0^*)} \ |W(\Sigma_0)|^2$.
 We note now that unlike the case of global supersymmetry the potential is no longer degenerate for the
 three vacua~\cite{Weinberg:1982id,Chamseddine:1982jx}.
  Suppose we add a term to  the superpotential and make the vacuum energy for one case
 vanish. In that case it is easy to check that the vacuum energy for other cases will be negative, i.e., that the
 vacua will be anti-deSitter. This would imply that the  Minkowskian vacuum would be unstable in each
 case.  However, it turns out that vacuum stability is helped by gravity ~\cite{Coleman:1980aw}.
  In fact it has been shown that the Minkowski vacuum will be stable against any finite size perturbations
  ~\cite{Weinberg:1982id}. It should be noted that vacuum degeneracy is not lifted in all cases
  when gauge symmetry breaks even in the presence of gravity.

We turn now to the breaking of supersymmetry.
In supergravity grand unification the
 breaking of supersymmetry can be generated by a superhiggs potential similar to the breaking of
a gauge symmetry by a Higgs potential. A general form of the  superpotential for the superhiggs is given by
$W_{SH}(Z) = m^2 \kappa^{-1} f_s(\kappa Z)$ where the  function $f_s(\kappa Z)$ depends on the dimensionless
product $\kappa Z$.
Here the breaking of supersymmetry gives $ <Z>\sim O(\kappa^{-1})$ and
 $f_s(\kappa Z)\sim O(1)$. The gravitino mass in this case is $m_{3/2}\sim O(\kappa m^2)$.
 If we take $m$ to be intermediate scale of size $10^{10-12}$ GeV, then $m_{3/2}$ lies in the range
 $1-100$ TeV.  In early works a simple choice for $W_{SH}$ was made, i.e.,  $W_{SH}= m^2 (z+B)$.
The VEV of the field $Z$ which breaks supersymmetry has no direct interaction with the visible sector
since  $<Z>\sim \kappa^{-1}$ and thus any direct interaction between the superhiggs field $Z$ and
the visible sector fields would generate a   mass for the visible sector fields which would be $O(\kappa^{-1})$.

  To shield the visible sector from such large mass growths
 the breaking of supersymmetry is communicated to the visible sector by gravity mediation.
 A simple way to see this is to write the superpotential including superhiggs and matter fields so that~\cite{Chamseddine:1982jx,Barbieri:1982eh}
  $W_{\rm tot}  = W(Z_a) + W_{SH}(Z)$, where $W(Z_a)$ depends on the matter fields $Z_a$ and
 $W_{SH}(Z)$ depends only on the superhiggs field  $Z$ which breaks supersymmetry. The two
 are, however, connected via the supergravity scalar potential~\cite{Chamseddine:1982jx,Cremmer:1982en}
 , i.e.,
$V = e^{{\kappa^2} K}~[({K}^{-1})_j^i (\frac{\partial W}{\partial z_i}~+ ~\kappa^2 K_{,i} W)
(\frac{\partial W}{\partial z_i}~+ ~\kappa^2 K_{,i} W)^{\dagger}
 - 3 \kappa^2 |W|^2]~+ V_D$,  where $V_D$ is the D-term potential. 
 Integrating out the superhiggs field, one finds that the low energy theory in the visible sector does contains
  soft breaking generated by gravity mediation which are size  $m_s\sim  \kappa m^2$ and thus the soft
  breaking is free of the Planck mass.
 However,  supergravity models with grand unification contain another heavy mass, i.e., the GUT mass $M_G$ in addition to
 the Planck mass.  Including $M_G$ in the analysis in a grand unified supergravity model brings in another type of
 hierarchy problem, i.e.,
 one would have mass terms of the form
$m_s M_G, ~\alpha m_s M_G \cdots,  \alpha^n m_s M_G$, where $\alpha\equiv  (\kappa M_G)$. A remarkable aspect of supergravity unified models is that
 all items of the type above cancel or vanish~\cite{Chamseddine:1982jx}.
In spontaneously broken supergravity the sum rule for masses  is given by ~\cite{Cremmer:1978iv,Nath:1983fp}
$\sum_{J=0}^{3/2} (-1)^{2J} (2J+1) m_J^2 = 2(N-1) m_{3/2}^2$,
where $N$ is the number of chiral superfields.

  In the universal supergravity models the effective low energy theory after integration over the heavy fields is thus
   given by~\cite{Chamseddine:1982jx}
$V_{eff} = |\frac{\partial \tilde W}{\partial Z_{\alpha}}|^2  +
m_0^2 Z_{\alpha}^{\dagger} Z_{\alpha} + (B_0 W^{(2)} + A_0 W^{(3)} + h.c)  + D$-term,
where $W^{(2)}$ and  $W^{3)}$ are parts of the superpotential which are bilinear and trilinear in the
scalar fields. Here the soft parameters consist of the universal scalar mass $m_0$, and 
the dimensioned parameters $A_0$  and $B_0$.  In the presence of non-universalities, one may have flavor dependence on
 the scalar masses as well as on the trilinear couplings. In the analysis above the vacuum energy has been
 adjusted to vanish. We note in passing that there exist models in which this is not the case and the minimum has
 a large negative energy~\cite{Ibanez:1982ee}.
 The class of supergravity unification is rather large, as it encompasses a vast variety of  particle physics models depending
 on the choice of the Kahler potential, the superpotential and the gauge kinetic energy function.
 Further,  strings in low energy limit are describable by supergravity models
 (for a review see ~\cite{Nath:2016qzm}).
 Similarly M-theory~\cite{Witten:1995ex} in low energy limit can be described by 11D supergravity~\cite{Cremmer:1978km} which can  be reduced further to 4D. Some examples of low energy string based supergravity models are
 KKLT~\cite{Kachru:2003aw} and the Large Volume Scenario~\cite{Balasubramanian:2005zx}.

    Soft terms also arise for gaugino masses. One such source is via loops  through their gauge interactions
  with the fields in the heavy sector~\cite{AlvarezGaume:1983gj,Nath:1983iz}. Thus the  gaugino masses
 $\tilde m_i, i=1,2,3$  for the   $U(1), SU(2), SU(3)$ gauge groups are given by
$\tilde m_i = \frac{\alpha_i}{4\pi} m_{3/2} C {D(R)}/{D(A)}$,
where  $D(R)$ is the dimensionality of the exchanged representation, $D(A)$ is the dimensionality of the
  adjoint representation and $C$ is the quadratic Casimir of the fields that contribute.
   One can also generate gaugino masses at the tree level via a field dependent gauge kinetic energy function
   $f_{\alpha\beta}$
    after   spontaneous breaking of supersymmetry and one has
 $\cl_{m}^{\lambda} = -(1/2) m_{\alpha\beta}  \bar \lambda^\alpha \lambda^\beta$.
Here $f_{\alpha\beta}$ transforms like the symmetric product of adjoint representations.
For the case when $f_{\alpha\beta}$ transforms like a singlet of the gauge group
$ \cl_{m}^{\lambda} = -\frac{1}{2} m_{1/2}  \bar \lambda \lambda $.  In general, however, one would have
non-universality of gaugino masses when $f_{\alpha\beta}$ does not transform as a pure singlet
\cite{Ellis:1985jn,nonuni2,Martin:2009ad,Feldman:2009zc}.
SUSY breaking can also arise  from  gaugino condensation such that $<\lambda \gamma^0 \lambda> \neq  0$~\cite{Nilles:1982ik}. This kind of breaking is often used in string model building.  The Higgs boson mass measurement
at $125$ GeV gives strong support to supergravity models with gravity breaking (see, e.g., ~\cite{Arbey:2012dq}).

Astrophysical evidence suggests the existence of dark matter in great abundance in our universe.
   Thus up to 95\% of the universe may be constituted of dark matter or dark energy. Here we will focus on
   dark matter.
Some leading candidates for dark matter include the
 weakly interacting massive particles (WIMPs), the extra weakly interacting particles~\cite{Feldman:2006wd}, and axions
 among many others. In supergravity models several neutral particles exist which could be possible candidates for dark matter, such as the neutralino, the sneutrino,  and the gravitino~\cite{Pagels:1981ke,Weinberg:1982zq}. We will focus here on neutralino type WIMP which is an odd R parity particle.  It has been shown that such a particle turns out to be the lightest supersymmetric particle in a large part of the parameter space of supergravity models~\cite{Arnowitt:1992aq} and if R -parity is conserved, it becomes a candidate for dark matter.
Indeed the neutralino was proposed as a possible candidate for dark matter soon after the formulation of
supergravity grand unified models~\cite{Goldberg:1983nd}.

The relic density of dark matter turns out to be an important constraint on model building.
 In supergravity grand unified models under the constraint that the weak SUSY scale is high perhaps lying
 in the several
    TeV region, the neutralino turns out to be most often a bino.  The annihilation cross section for bino-like  neutralino is rather small which implies that the neutralinos cannot efficiently annihilate to produce the desired relic density consistent with the data from WMAP~\cite{Larson:2010gs}   and from PLANCK~\cite{Ade:2015xua}
    experiments.  In this case we need
coannihilation~\cite{Griest:1990kh,Nath:1992ty,Bell:2013wua,Baker:2015qna}.
 Coannihilation is a process in which one or more sparticles other than the neutralino enter in the annihilation process modifying the  Boltzman equation that controls the relic density.
 Analysis shows that if the next to the lightest supersymmetric  particle (NLSP) lies close to the LSP in mass then there can be a significant enhancement in the annihilation of the LSP allowing one to satisfy the relic density constraints consistent with the WMAP and the PLANCK data.  {Coannihilation, however, makes the detection of supersymmetric
 signals more difficult because the decay of the NLSP leads to soft final states which often do not pass the
 kinematical cuts for the conventional signal regions. }

Cosmological models  with cold dark matter such as $\Lambda$CDM have been  pretty successful in explaining  the large-scale structure of the universe. At small scales, however,  some features arise which require
attention~\cite{Weinberg:2013aya}.
The first of these is the
cusp-core problem.
 This relates to the fact that the observed galaxy rotation curves are better fit by constant dark matter density cores
 which is the Berkert profile,
 $\rho ( r) =  { \rho_0 r_0^3}/{ [(r + r_0) (r^2 + r_0^2)]}$
 rather than the NFW profile $\rho ( r) =  { \rho_0 \delta }/{ [\frac{r}{r_s} (1+ \frac{r}{r_s})^2]}$ which is produced
in N-body simulations  using CDM.
The second is the so called ``missing satellite'' problem where  CDM predicts too many dwarf galaxies.
More detailed analyses indicate that both problems could be
solved by inclusion of complex dynamics and baryonic physics~\cite{Governato}.
 Aside from complex dynamics and baryonic matter,
another possibility  involves particle physics models such as  (repulsive) self-interactions.
or a  dark particle of de Broglie wavelength of
$\sim 1\kpc$ and a mass which lies in the range of $10^{-21}-10^{-22}$ eV~\cite{Kim:2015yna}.
}

\section{Unification in strings \label{sec7}}

{
  As discussed in section 6,
 N=1 supergravity grand unified models with a hidden sector lead to the breaking of supersymmetry in a phenomenologically
viable manner and with  three generations of quarks and leptons  allow
an extrapolation of physics from the electroweak scale to the grand unification scale. Above this scale one expects
 quantum gravity effects to become important. The next step then is to look for a theory of quantum gravity
whose low energy limit is supergravity grand unification.
Possible candidates are string theories which come in several varieties: Type I, Type IIA, Type IIB, heterotic $SO(32)$ and heterotic $E_8\otimes E_8$ \cite{gross}  .
The Type I and Type II strings contain $D$ branes. The D branes
can support gauge groups and chiral matter can exist at the intersection of D branes.  The
various string Types can arise from the so called M-theory whose low energy limit is 11D
supergravity. Most of the model building in string theory has occurred in heterotic
$E_8\otimes E_8$/Horava -Witten theory~\cite{Horava:1995qa}, where the Horava-Witten theory 
arising  from the  low energy limit of M theory on  $R^{10}\times S^1/Z_2$  may be viewed as the strong coupling limit of $E_8\otimes E_8$,
and  Type IIB/F theory, where F theory~\cite{Vafa:1996xn} may be viewed as the strong coupling limit of Type IIB strings.
One can also compactify M-theory on other manifolds such as on a manifold $X$ of $G_2$ holonomy~\cite{Duff:2002rw,Acharya:2001gy,Acharya:2015oea}.
One  problem in working with string models is that they possess a huge number of vacuum states~\cite{Douglas:2006es}, as many as
$10^{500}$, and the search for the right vacuum state that describes our universe is a daunting task.

One possible way to proceed  then is to thin out the landscape of vacuum states  by
imposing phenomenological constraints. The most obvious ones are the emergence of  groups
which support chiral matter which limits the groups to $SU(N), SO(4N+2), N\geq 1$ and  $E_6$ after reduction to four dimensions,
an $N=1$ supersymmetry in 4D,  a hidden sector that allows for breaking of supersymmetry, and
three generation of chiral fermions  which correspond to the three observed  generations  of quarks and leptons.
 For the emergence of supergravity grand unified model
it is also of relevance that the grand unification scale $M_G$ emerge in some natural way from the Planck
scale where $M_G$
is expected to be close to the scale where the 10-dimensional theory reduces to four dimensions.

 The $E_8\otimes E_8$ heterotic string
 was one of the first to be investigated at significant length~\cite{Candelas:1985en}.
  After compactification to four dimensions it has a rank which can be up to 22. This allows for many possibilities for model building which have been pursued in the context of using
       free field constructions, orbifolds, and  Calabi-Yau compactification
       (for a sample of the early phenomenological analyses see \cite{early-string-pheno}).
        Most of the analyses are within Kac-Moody level 1.  Here one can achieve unified groups such as~\cite{lewellen}
        $SU(5), ~SO(10)$ or $E(6)$. However, one does not have scalars in the adjoint representation to break the
        gauge symmetry.  At level 2, adjoint scalars are achievable but three massless generations are not easy to get.
        At level 3, it is possible to have scalars in the adjoint representation and also 3 massless generations.
          However, here one finds that the quark-lepton textures are rank 3 and thus difficult to get realistic quark-lepton
          masses~\cite{Kakushadze:1997mc}.
                   In addition to the heterotic string constructions, a large number of string model constructions have since appeared, and a significant
     number within Type II (for reviews see \cite{Angelantonj:2002ct,Ott:2003yv,Blumenhagen:2005mu,Blumenhagen:2006ci}).           Of course in such models which have
an effective theory with $N=1$
supersymmetry, one would need to break supersymmetry to make
contact with observable physics~(see, e.g., \cite{Kors:2003wf}).

  We note in passing that  while it is desirable  that grand unification arise from strings (for a recent review
  see ~\cite{Raby:2011jt}) it is not essential that it do so   since string theory is already a unified
   theory and  it  is not necessary for us to insist on grand unification in the effective low energy theory.
   Rather we may have the standard model gauge
   group emerging  directly from string theory without
   going through grand unification. In this case we will have
$g_i^2k_i=g_{string}^2={8\pi G_N}/{\alpha'}$,
	 where  $G_N$  is the Newtonian constant, $\alpha'$  is the
	Regge slope, and $k_i$ are the Kac-Moody levels of  the subgroups.
	There are positive integers for non-abelian gauge groups\cite{Schellekens:1989qb} while for $U(1)$
        the normalization of $k$ is arbitrary.

	   Of course string theory is supposed to unify gravity along with other forces, and
	   one may look at the evolution of the dimensionless parameter $\alpha_{\rm GR} = G_N E^2$
	  along with the other three couplings (for a review see~\cite{Dienes:1996du}).
	 	   In the  normal evolution of the gauge and gravity couplings using the spectrum of MSSM,
	   one finds that while the gauge couplings do unify at the scale $M_G\sim 10^{16}$ GeV,
	   the gravity coupling $\alpha_{\rm GR}$ does not. There are several possible ways to address
	  this lack of unification. One possibility is that as we evolve the couplings above the
	   scale $M_G$,  $\alpha_G$
	   and $\alpha_{\rm GR}$  will unify.
	   Another possible way is  that at some scale
	   below $M_G$,  a new dimension of space opens up. Here if matter resides on the 4 dimensional
	   wall while gravity resides in the bulk,  the evolution of $\alpha_{\rm GR}$ will be much rapid and
	   there may be unification of gauge and gravity couplings at a scale much below the Planck scale.
   A  further modification of this idea is strings where the string scale is lowered to lie at the TeV scale.
   In this case gravity can get strong at a much lower scale and unification of gauge and gravity coupling
   can occur in the few TeV region~\cite{Lykken:1996fj,Arkani-Hamed:1998rs,Antoniadis:1998ig}.
   More recent progress in model building has come from F-theory which as noted earlier can be
   viewed as a strong coupling limit of Type IIB string. For some recent work in F-theory model building
   see, e.g.~\cite{Beasley:2008dc,Donagi:2008kj,Dolan:2011aq,Grassi:2014zxa,Callaghan:2013kaa}.
   }

\section{Monopoles\label{sec8}}
{

It was shown by Dirac~\cite{Dirac:1948um} that in Maxwell electrodynamics the existence of a monopole of magnetic charge $g$
would imply a quantization of the electric charge so that $e.g.=\frac{1}{2} n \hbar$.  In unified theories while
$SU(2)\times U(1)$ does not possess a magnetic monopole, $SO(3)$ does and it arises as a consequence of solution to the
field equations. The quantization condition in this case is $e.g.= n\hbar$ which is  the Schwinger quantization~\cite{Schwinger:1966nj}.
Grand unified theories  also possess  monopoles and they appear again as solutions to field equations.
Unlike the Maxwell electrodynamics where the monopole may or may not exist, in grand unified theories
where  a $U(1)$ arises as a reduced symmetry, the monopole is a necessity and a prediction.
 However, the monopoles in grand unified theories will be superheavy
with a mass of size the GUT scale.
One problem associated with these monopoles is that monopoles produced in the early universe
 would contribute a matter density in excess of the critical relic
density which would over close the universe. Inflationary cosmology solves this problem.
In some models the magnetic monopoles can be much lighter. Monopoles also appear in
intersecting D-brane models where they appear along with color singlet fractionally charged states~\cite{Kephart:2017esj}.
For the current experimental status of magnetic monopoles see~\cite{Milton:2006cp}.

 }

\section{Proton decay in GUTs and  Strings\label{sec9}}
{

One of the hallmarks of most unified models  is the prediction  that proton will decay.
It is also a possible discriminant of models based on GUTs vs strings.
One of the predictions in both supersymmetric and non-supersymmetric grand unification
is the  proton decay mode $p\to e^+ \pi^0$. This  mode arises from
 dimension six operators and involves the exchange of  leptoquarks
(for reviews see Refs. \cite{Nath:2006ut,Raby:2008pd,Hewett:2012ns}). A rough estimate of the decay width is
 $\Gamma (p \rightarrow e^+ \pi^0) \simeq  \alpha_G^2 \frac{m_p^5}{M_V^4}$ where $\alpha_G=g_G^2/4\pi$
 with $g_G$ being the unified coupling constant, and $M_V$ the lepto-quark mass. It leads to a partial lifetime
 of  $\tau (p \rightarrow e^+ \pi^0) \simeq 10^{36\pm1} yrs$.
This mode allows the possibility of discrimination among models based on GUTs  vs those based on strings.
Thus a generic  analysis of D brane models allows only
$10^2\cdot\overline{10}^2$ SU(5) type dimension six operators which leads to  the decay
$p\to \pi^0 e^+_L$\cite{Klebanov:2003my}.
 In $SU(5)$ grand unification, one has in addition the operator
$10\cdot\overline{10}\cdot5\cdot\overline{5}$ which allows $p\to \pi^0e^+_R$. Further, generic D brane models do not allow  the decay
$p\to \pi^+\nu$ while $SU(5)$ grand unification does.
It has been pointed out, however, that special regions of intersecting D brane models exist  which
allow the operator $10\cdot\overline{10}\cdot5\cdot\overline{5}$ and the  purely stringy proton decay rate can be of the same order as the one from SU(5) GUTs  including the mode $p\to \pi^+\nu$~\cite{Cvetic:2006iz}.
The current experimental status of proton decay for this partial decay mode is the following:
  Superkamiokande gives the limit
 $\tau(p\to e^+\pi^0) >2\times 10^{34} ~{\rm yrs}$~\cite{TheSuper-Kamiokande:2017tit}
 while in the future
 Hyper-K is expected to achieve a  sensitivity of
   $\tau(p\to e^+\pi^0) > 1\times 10^{35}~{\rm yrs}$~\cite{Abe:2011ts}.

   In supersymmetric unified models there are additional operators that can generate proton decay.
  Thus if R parity is not conserved,  proton decay can proceed with baryon and lepton number violating dimension
   four operators. Such  a decay is too rapid and must be eliminated by imposition of R parity conservation.
   In this case we still have baryon and lepton number violating dimension five operators
      arising from the exchange of Higgs triplets,  which give rise to
   proton decay~\cite{dim5,Nath:2006ut}.
      Here  the dominant decay mode is $p\to K^+\bar \nu$ and could also be dangerous in terms of
   proton stability~\cite{Dimopoulos:1981dw,Ellis:1981tv,Nath:1985ub,Hisano:1992jj,Lucas:1996bc,Babu:1998wi,Goto:1998qg,Bajc:2002bv}.
   The SUSY decay modes depend  sensitively  on the sparticle spectrum as well as on
  CP phases~\cite{Ibrahim:2007fb,Ibrahim:2000tx}.  
  The current experimental limit from Super-Kamiokande  is $\tau(p\to \bar \nu K^+) > 4\times 10^{33} {\rm yrs}$, while
in the future it is expected that Hyper-K may reach a sensitivity of
$\tau(p\to \bar \nu K^+) > 2\times 10^{34} ~{\rm yrs}$.

It is known that
 proton decay lifetime from  baryon and lepton number violating dimension five operators in SUSY GUTs with a low sparticle spectrum would lie below the current experimental limits.
One possibility for stabilizing the proton is via a  cancellation mechanism~\cite{Nath:2006ut,Nath:2007eg}.
The other possibility is via using a heavy sparticle spectrum which enters in the loop diagrams.
Very roughly the proton decay from dimension five operators has the sparticle mass dependence of
 $m_{\chi^{\pm}_1}^2/m_{\tilde q}^4$ where $\chi_1^{\pm}$ is the chargino and the $\tilde q$ is the squark.
 This means that a  suppression of proton decay can be achieved with a large sfermion mass.
 As  will be discussed  in section~\ref{sec10}, the
 discovery of the Higgs boson with a  mass of $\sim 125$ GeV points to a high SUSY scale and a high
 SUSY scale implies a larger proton decay lifetime for the SUSY modes. In fact one finds  a direct correlation between
 the proton lifetime and the Higgs boson mass\cite{Liu:2013ula} which shows that the experimental lower limit
 on the proton lifetime for the SUSY mode can be easily met under the Higgs boson mass constraint.
}

\section{A stronger case for SUSY/SUGRA after Higgs boson discovery\label{sec10}}

{
The measurement of the Higgs  boson at $125$ GeV gives further support for supersymmetry.
One reason for that is vacuum stability. For large field configurations where $h>> v$ the Higgs potential is governed by the quartic term $V_h\sim \frac{\lambda_{eff}}{4} h^4$.
For vacuum stability $\lambda_{eff}$ must be positive. In the
  Standard Model  analyses  done using NLO and NNLO corrections
show that  with a 125 GeV Higgs boson, the vacuum can be stable only up to
about $10^{10}-10^{11}$ GeV~\cite{Degrassi:2012ry}.
For vacuum
  stability up to the Planck scale one needs $m_h> 129.4\pm 1.8$ GeV.
 Vacuum stability depends critically on the top mass.  A larger top mass makes the vacuum more
unstable.
An advanced precision analysis~\cite{Bednyakov:2015sca}
   including two-loop matching, three-loop renormalization group evolution, and pure QCD corrections through four loops
   gives an upper bound on the top pole mass for the stability of the standard model vacuum
   up to the Planck mass scale
   of $m_t^{\rm cri} = (171.54 \pm 0.30^{+0.26}_{-0.41}) {\rm ~GeV}$,
   while the experimental value of the top is $m^{\rm exp}_t= (173.21\pm 0.51 \pm0.71) {\rm ~GeV}$.
Though an improvement on previous analyses the stability of the \sm vacuum is still not fully guaranteed.
In models based on supersymmetry with a Higgs mass of $125$ GeV, the  vacuum is stable up
to the Planck scale. Of course one has to choose the parameter space of supergravity models appropriately.

As discussed in sec \ref{sec6} the Higgs boson mass measurement at $125$ GeV~\cite{Chatrchyan:2012ufa,Aad:2012tfa}
indicates that the loop correction to the Higgs boson mass in
supersymmetry is large. This in turn indicates that the weak SUSY scale, and specifically the scalar masses,
 must be large lying in the several TeV region~\cite{Akula:2011aa,Arbey:2012dq,susy-higgs,Baer:2015fsa}.
 It turns out that large scalar masses arise naturally in
 the radiative breaking of supergravity.
  Thus the radiative breaking in supergravity unified models obeys the
 relation (for a review see~\cite{Ibanez:2007pf})
 $M_Z^2+  2\mu^2 \simeq {(1-3 D)}m_0^2 + C (m_0, A_0, \tan\beta)~~C>0$,
where $C$ is a polynomial in $m_0$ and $A_0$ and
  $D$ depends on loop corrections, Yukawas and the weak SUSY scale.
For the case when  $D<1/3$ there is an upper bound on sparticles masses for fixed $\mu$.
This the case of ellipsoidal geometry (EB) where for a given  $\mu$, one has an upper limit
on how large  the soft parameters can get.  When
   $D\geq 1/3$  the geometry becomes hyperbolic (HB) and for a fixed $\mu$ the scalar masses
   get large
   ~\cite{Chan:1997bi}.

      The HB contains focal curves and focal surfaces (see Akula etal.  in \cite{Chan:1997bi}).
 The transition point between the two branches, i.e., between EB and HB, is  $D=1/3$ referred to as the focal point.
 Here $\mu$ essentially becomes independent of $m_0$. Thus in general on the focal point, focal curves and focal
 surfaces,  $\mu$ can remain small while scalar masses get large.
On HB the gauginos  can be light and discoverable at the LHC even if the SUSY scale $M_S$ is in the
 TeV region. Further, the weak SUSY scale is approximated by $M_S= \sqrt{m_{\tilde t_1} m_{\tilde t_2}}.$
 where $\tilde t_1, \tilde t_2$ are the stop masses.
Thus $m_{\tilde t_1}$ may be a few hundred GeV  while
$m_{\tilde t_2}$ is order several TeV which leads to $M_S$ in the TeV region~\cite{Kaufman:2015nda}.
This means that even for the case when the weak SUSY scale is large, one of the stops could have a mass that lies
in the few hundred GeV region and may be discoverable. Of course, for non-universal supergravity  models
the  sleptons can be much lighter than the squarks. Here we note that in searches for supersymmetry
optimization of signal analysis beyond what is employed in simplified models 
 is  often necessary as important regions of the  parameter space can
otherwise be missed (see, e.g., \cite{Chatterjee:2012qt,Nath:2016kfp,Choudhury:2017lxb,Aboubrahim:2017aen}).

 It is possible, however, that the weak scale of SUSY is much larger than previously thought and could
 be upwards of 10 TeV.
 A weak SUSY scale of this size is consistent with radiative breaking of
 the electroweak symmetry,
the Higgs boson mass constraint and the relic density constraint~\cite{Aboubrahim:2017wjl}. It can also resolve
the so called gravitino decay problem in supergravity  and string theories.
Thus an
  intrinsic element of supergravity unified models is the existence of a gravitino which is a supersymmetric
partner of the graviton and becomes massive after spontaneous breaking of supersymmetry.  The gravitino
could be either stable or unstable. If it is stable it would be the lowest mass odd $R$ parity object and
thus contribute to dark matter. In this case one finds that the mass of the gravitino must be less than $1$KeV
in order that it not over close the universe. If the gravitino is not the LSP, it would be unstable and decay
and there are strong constraints on the gravitino in this case. Here one needs to make certain that the
gravitino which has only gravitational interactions does not decay too late, i.e., does not occur
 after the BBN time, i.e., $(1-10^2)$s.  which would upset  the successes of the BBN.
 As already noted
 one of the implications of the Higgs boson measurement, is that the  weak SUSY scale which is typically set by
 the gravitino mass is high lying in  the several TeV region.  However, the scalar masses though scaled by
 the gravitino mass
 need not be equal to the gravitino mass.

  We note that supergravity models  have a  large landscape
 of soft parameters~\cite{Francescone:2014pza}
   which  include non-universalities in the gaugino sector  and as well as non-universalities in
  the matter and Higgs sectors~\cite{NU}.
   First there could be non-universalities which are model dependent
 and split the scalar masses
 and second that renormalization group  evolution has significant effect on the scalar masses when evolved down to the electroweak
 scale.  Thus in general the scalar masses could be much lower than the gravitino mass.  Further, the gaugino
 masses  could be significantly smaller than the scalar masses. This would allow  the gravitino to have decay
 modes $\tilde{g}g, ~ \tilde{\chi}_1^{\pm}W^{\mp}, ~\tilde{\chi}_1^0\gamma,~\tilde{\chi}_1^0 Z$.
 { We note in passing that supergravity theories arising from  string compactification with
 stabilized moduli   often
lead to a gravitino mass which may lie in the range 10 TeV and above (see, e.g., ~\cite{Halverson:2017deq,Kane:2009kv,Ellis:2017vpy}).}

 For  the  gravitino mass lying in the 50-100 TeV region, the decay of the gravitino   occurs significantly
 before the BBN time~\cite{Aboubrahim:2017wjl}.  There is, however, one more constraint that one needs to attend to, which is that
 under the assumption of R parity conservation, each of the gravitino decay will result in an LSP neutralino
 which will contribute to the relic density of the universe. Thus  here the total relic density of the neutralinos
 will be given by $\Omega_{\tilde{\chi}^0_1} =  \Omega^{\rm th}_{\tilde{\chi}^0_1} +  \Omega^{\G}_{\tilde{\chi}^0_1}$
 where  $\Omega^{\rm th}_{\tilde{\chi}^0_1}$ is the regular relic density arising from the
 thermal production of neutralinos  after the freeze out and  $\Omega^{\G}_{\tilde{\chi}^0_1}$
 is the relic density arising from the decay of the gravitinos. We may write $\Omega^{\G}_{\tilde{\chi}^0_1}$
 so that  $\Omega^{\rm \G}_{\tilde{\chi}^0_1} =  ({m_{\tilde{\chi}_1^0}}/{m_{\tilde G}})  \Omega_{\tilde G}$,
 which implies that we need to compute the quantity  $\Omega_{\tilde G}$.
 We assume that  the gravitinos produced in the early universe before the start of inflation have been
 inflated away and the relevant relic density of gravitinos is the one that is produced after inflation during
 the reheating period.

 Detailed analyses of this production exists in the literature and one finds that the relic density of neutralinos
 generated 
  by the decay of the gravitinos produced by reheating is given by
 $\Omega^{\G}_{\chi_1^0}h^2$
        =  $ \sum_{i=1}^{3}
        \omega_i\, g_i^2$
       $ \left(1+{m_i^2}/{3m_{\tilde G}^2}\right)$
        $\ln\left({k_i}/{g_i}\right)
        \left({m_{\tilde{\chi}_1^0}}/{100 \GeV}\right)
        \left({T_R}/{10^{10} \GeV}\right)$.
        Here $g_i, m_i (i=1,2,3)$ are the gauge couplings and the gaugino masses for the gauge groups
        $U(1)_Y$, $SU(2)_L$ and $SU(3)_C$ which are evaluated at the reheat value $T_R$ using
       renormalization group  to evolve their values from the GUT scale.  Further,  $\omega_i(i=1,2,3)= (0.018, 0.044,    0.177)$~\cite{Pradler:2006qh}.
        Analysis shows that negligible amount of the relic density arises from the decay of the gravitino
        up to reheat temperatures of $10^{10}$ GeV~\cite{Aboubrahim:2017wjl}.

}

\section{Testing supergravity unification at future colliders\label{sec11}}
{
As mentioned in section 10, the case for SUSY/SUGRA is much stronger as a consequence of the Higgs boson discovery and the measurement of its mass at 125 GeV.  Currently there
 is no other paradigm that can  replace supergravity grand unification as we extrapolate physics from the electroweak scale to the  GUT scale. For these reasons the search for
sparticles must continue.  There is a good chance that we will find sparticles at the LHC by the time all its runs are over.
The discovery of even one sparticle will open up a new era for sparticle spectroscopy including the discovery
of the remaining sparticles, and  precision measurement of their masses and couplings.  For these higher energy colliders
are essential.
For the future several proposals are under consideration both for high energy $e^+e^-$ colliders
as well as for high energy proton-proton colliders. For the  $e^+e^-$ colliders the possibilities are:
(a) ILC:  International Linear Collider, (b) CEPC: Circular  Collider, (c) FCC-ee (TLEP): Future  Circular  Collider.
The ILC is under consideration in Japan, CEPC in China and FCC-ee at CERN.
These colliders are essentially  Higgs factories which are likely to run  at an energy around $240$ GeV which gives the optimal cross section for Higgsstrahlung, i.e.,  $e^+e^-\to Zh$.  The $Zh$ final state is the preferred mode rather than $hh$ since $Z$ can be efficiently detected via  $Z\to \ell^+\ell^-$. The Higgs factories can
do precision physics related to the Higgs boson specifically the couplings of the Higgs  bosons to fermions
and other  electroweak parameters with great accuracy. Some supersymmetric effects could show up in
these high precision experiments.
The sparticles most likely to be probed would be electroweak
gauginos and light sleptons if they are low mass.  For the study for most other sparticles one needs
 proton-proton colliders.  Here  the possibilities are (i) SppC:
 70-100 TeV  pp collider, (ii)  VLHC: 100 TeV pp collider.
The SppC if built would be in China and VLHC is a possibility for the future at CERN.
}

\section{Conclusion\label{sec12}}
{
There are a variety of reasons why supersymmetry is  desirable when we think of high scale physics.
  One reason is the well
 known hierarchy problem~\cite{Gildener:1976ih}
    arising from loop corrections to the scalar masses.     Thus while a loop correction to a
  fermion mass is proportional to the fermion mass, i.e.,  $\delta m_f \propto m_f$,
 for the scalars  the corrections  to  the scalar mass $m$  is of  the form  $ \delta m^2 \propto \Lambda^2$,
  where $\Lambda$ is a cutoff scale. In a quantum field theory the cutoff scale could be order the
  Planck mass and thus the correction is very large.
  One of the ways to overcome this problem is supersymmetry where the loop correction from
the squark  exchange cancels the loop correction from the quark exchange.  This leads to a natural cancellation
of  1 part in $10^{28}$.
The situation is similar to the $\Delta S=1$ neutral current case in the \sm
where the charm quark loop cancels the $u$ quark loop consistent with experiment
 $Br(K^0\to \mu^+\mu^-)/Br(K^+\to \mu^+ \nu_\mu)=   (6.84\pm 0.11) \times 10^{-9}$~\cite{Patrignani:2016xqp}.
In this case the cancellation required  is order one part in $10^{9}$ and it leads to the
discovery of the charm quark.
In comparison for the case of the Higgs boson mass, the cancellation is  1 part
in $10^{28}$ between the squark loop and the quark loop and thus there is an overwhelming  reason for supersymmetry to be discovered.

In addition to  the above let us recount some of the other successes of the supersymmetry/supergravity models.
 One of the early ones includes the fact that supersymmetry/supergravity models provide
 the right number of extra particles  needed for the unification of couplings using the LEP data as one
 moves from the electroweak scale to the grand unification scale.
Supergravity models provide an explanation for
the tachyonic Higgs mass term that is central to accomplish spontaneous breaking of the electroweak
symmetry in the Standard Model.
Supergravity grand unification predicted the Higgs boson mass to lie below 130 GeV which is consistent
with the current measurement of the Higgs boson.
Currently there is no good alternative to  supergravity grand unification  if we wish to
  extrapolate physics from the electroweak scale to the grand unification scale.
 So the search for sparticles must continue. The most likely sparticle candidates for discovery are the
 neutralino $\tilde \chi^0$,  the chargino $\tilde \chi^{\pm}$, the gluino $\tilde g$, the stop
 $\tilde t_1$, and the stau $\tilde \tau_1$. There is also the possibility of discovering the additional
 Higgs bosons that appear in SUSY extensions of the standard model.
   There is a good chance that with  the full capability of the LHC ($\cl =3000$ fb$^{-1}$, $\sqrt s=14$ TeV)   one discovers sparticles and/or additional Higgs boson. At the same
 time a good idea to look ahead and plan for a 100 TeV pp super collider.\\

}

\noindent
\textbf{Acknowledgments: }
This research was supported in part  by  NSF Grant PHY-1620526.\\


\small{

}

\begin{thebibliography}{999}

\bibitem{Glashow:1961tr}
  S.~L.~Glashow,
  Nucl.\ Phys.\  {\bf 22}, 579 (1961);
S.~Weinberg,
  Phys.\ Rev.\ Lett.\  {\bf 19} (1967) 1264;
A. Salam, in Elementary Particle Theory, ed. N. Svartholm
(Almquist and Wiksells, Stockholm, 1969) p.367;
  G.~'t Hooft,
  Nucl.\ Phys.\ B {\bf 35}, 167 (1971). See also
  Nucl.\ Phys.\ B {\bf 33}, 173 (1971);
  G.~'t Hooft and M.~J.~G.~Veltman,
  Nucl.\ Phys.\ B {\bf 44}, 189 (1972);
 H.~D.~Politzer,
  Phys.\ Rev.\ Lett.\  {\bf 30} (1973) 1346;
  D.~J.~Gross and F.~Wilczek,
  Phys.\ Rev.\ Lett.\  {\bf 30} (1973) 1343.


\bibitem{Pati:1974yy}
  J.~C.~Pati and A.~Salam,
  Phys.\ Rev.\ D {\bf 10}, 275 (1974)
  Erratum: [Phys.\ Rev.\ D {\bf 11}, 703 (1975)].

\bibitem{Pati:1973uk} 
  J.~C.~Pati and A.~Salam,
  Phys.\ Rev.\ D {\bf 8}, 1240 (1973).



\bibitem{Georgi:1974sy}
  H.~Georgi and S.~L.~Glashow,
  Phys.\ Rev.\ Lett.\  {\bf 32}, 438 (1974).
 


\bibitem{georgi-so10}
H. Georgi, in \textit{Particles and Fields}, ed. C. E. Carlson (AIP,
New York, 1975) 575;
H.~Fritzsch and P.~Minkowski, Annals Phys.\  {\bf 93} (1975) 193.



\bibitem{Gildener:1976ih}
  E.~Gildener and S.~Weinberg,
  Phys.\ Rev.\ D {\bf 13}, 3333 (1976).


\bibitem{Wess:1974tw}
  J.~Wess and B.~Zumino,
  Nucl.\ Phys.\ B {\bf 70} (1974) 39;
 J.~Wess and B.~Zumino,
  Phys.\ Lett.\ B {\bf 49} (1974) 52;
 D.~V.~Volkov and V.~P.~Akulov,
  JETP Lett.\  {\bf 16} (1972) 438.


\bibitem{Nath:1975nj}
  P.~Nath and R.~L.~Arnowitt,
  Phys.\ Lett.\  {\bf 56B}, 177 (1975).
  %
\bibitem{Arnowitt:1975xg}
  R.~L.~Arnowitt, P.~Nath and B.~Zumino,
  Phys.\ Lett.\  {\bf 56B}, 81 (1975).
  



\bibitem{Freedman:1976xh}
  D.~Z.~Freedman, P.~van Nieuwenhuizen and S.~Ferrara,
  Phys.\ Rev.\ D {\bf 13}, 3214 (1976);
  S.~Deser and B.~Zumino,
  Phys.\ Lett.\  {\bf 62B}, 335 (1976).


\bibitem{Nath:2016qzm}
  P.~Nath,
  ``Supersymmetry, Supergravity, and Unification,''
   Cambridge, Uk: Univ. Pr. (2016) 520 P. (Cambridge Monographs On Mathematical Physics).


\bibitem{Chamseddine:1982jx}
  A.~H.~Chamseddine, R.~L.~Arnowitt and P.~Nath,
  Phys.\ Rev.\ Lett.\  {\bf 49}, 970 (1982);
  P.~Nath, R.~L.~Arnowitt and A.~H.~Chamseddine,
  Nucl.\ Phys.\ B {\bf 227}, 121 (1983);
  L.~J.~Hall, J.~D.~Lykken and S.~Weinberg,
  Phys.\ Rev.\ D {\bf 27}, 2359 (1983).

\bibitem{Georgi:1974yf} 
  H.~Georgi, H.~R.~Quinn and S.~Weinberg,
  Phys.\ Rev.\ Lett.\  {\bf 33}, 451 (1974).
  doi:10.1103/PhysRevLett.33.451


\bibitem{g1}
S.~Dimopoulos, S.~Raby and F.~Wilczek,
Phys.\ Rev.\ D {\bf 24}, 1681 (1981);.

\bibitem{g2}
J. Ellis, S. Kelley and D. V. Nanopoulos, Phys.
Lett. {\bf 249B}, 441,(1990)

\bibitem{g3}
U. Amaldi, W. de Boer and H. Furstenau, Phys.
Lett. {\bf 260B}, 447 (1991).


\bibitem{g4}
 P.~Langacker and M.~x.~Luo,
Phys.\ Rev.\ D {\bf 44}, 817 (1991).




\bibitem{Hill:1983xh}
  C.~T.~Hill,
  Phys.\ Lett.\  {\bf 135B}, 47 (1984);
  Q.~Shafi and C.~Wetterich,
  Phys.\ Rev.\ Lett.\  {\bf 52}, 875 (1984).

\bibitem{g5}
L.~J.~Hall and U.~Sarid,
Phys.\ Rev.\ Lett.\  {\bf 70}, 2673 (1993).


\bibitem{g6}
T.~Dasgupta, P.~Mamales and P.~Nath,
Phys.\ Rev.\ D {\bf 52}, 5366 (1995).






\bibitem{Green:1987sp}
  M.~B.~Green, J.~H.~Schwarz and E.~Witten,
  ``Superstring Theory. Vol. 1: Introduction,''
  Cambridge, Uk: Univ. Pr. ( 1987) 469 P. (Cambridge Monographs On Mathematical Physics);
  ``Superstring Theory. Vol. 2: Loop Amplitudes, Anomalies And Phenomenology,''
  Cambridge, Uk: Univ. Pr. ( 1987) 596 P. (Cambridge Monographs On Mathematical Physics).


\bibitem{seesaw}
  P.~Minkowski,
  Phys.\ Lett.\ B {\bf 67} (1977) 421;
    M. Gell-Mann, P. Ramond and R. Slansky,
   in {\it Supergravity}, eds. P. van Nieuwenhuizen et al.,
   (North-Holland, 1979), p.~315;
   T. Yanagida,
  KEK
   Report~79-18, Tsukuba, 1979, p.~95;
  S.L. Glashow, in {\it Quarks and Leptons}, Carg\`ese, eds. M. L\'evy et al.,
(Plenum, 1980), p. 707;
  R.~N.~Mohapatra and G.~Senjanovi\'c,
  Phys.\ Rev.\ Lett.\  {\bf 44} (1980) 912.




\bibitem{Pati:2017ysg}
  J.~C.~Pati,
  Int.\ J.\ Mod.\ Phys.\ A {\bf 32}, no. 09, 1741013 (2017).
  


\bibitem{Dimopoulos:1981zb}
E. Witten,  B185, 513 (1981);
 S.~Dimopoulos and H.~Georgi,
  Nucl.\ Phys.\ B {\bf 193}, 150 (1981);
  N.~Sakai,
  Z.\ Phys.\ C {\bf 11}, 153 (1981).


\bibitem{Grinstein:1982um}
  B.~Grinstein,
  ``A Supersymmetric SU(5) Gauge Theory With No Gauge Hierarchy Problem,''
  Nucl.\ Phys.\ B {\bf 206} (1982) 387.

\bibitem{Masiero:1982fe}
  A.~Masiero, D.~V.~Nanopoulos, K.~Tamvakis and T.~Yanagida,
  ``Naturally Massless Higgs Doublets In Supersymmetric SU(5),''
  Phys.\ Lett.\ B {\bf 115} (1982) 380.

\bibitem{flipped}
 Barr, S~M,  Phys. \ Lett. \ B {\bf 112} (1982) 219;
Derendinger,  J~P, Kim , J~E, and Nanopoulos,  D~V. Phys\ Lett.\  B {\bf 139} (1984) 170.





\bibitem{so10-largereps}
 T.E. Clark, T.K. Kuo, and N. Nakagawa, Phys. lett. B 115, 26 (1982);
C.S. Aulakh and R.N. Mohapatra,
 Phys. Rev. D 28, 217 (1983);
C.S. Aulakh, B. Bajc, A. Melfo, G. Senjanovic and F. Vissani, Phys. Lett. B 588, 196 (2004);
C.S. Aulakh and S.K. Garg, Nucl. Phys. B 757, 47 (2006);
  Nucl.\ Phys.\ B {\bf 857}, 101 (2012).


\bibitem{Babu:2005gx}
  K.~S.~Babu, I.~Gogoladze, P.~Nath and R.~M.~Syed,
  Phys.\ Rev.\ D {\bf 72}, 095011 (2005).


\bibitem{Babu:2006rp}
  K.~S.~Babu, I.~Gogoladze, P.~Nath and R.~M.~Syed,
  Phys.\ Rev.\ D {\bf 74}, 075004 (2006).


\bibitem{Mohapatra:1979nn}
  R.~N.~Mohapatra and B.~Sakita,
  Phys.\ Rev.\ D {\bf 21}, 1062 (1980);
  F.~Wilczek and A.~Zee,
  Phys.\ Rev.\ D {\bf 25}, 553 (1982).


\bibitem{Nath:2001uw}
  P.~Nath and R.~M.~Syed,
  Phys.\ Lett.\ B {\bf 506}, 68 (2001).


\bibitem{Nath:2001yj}
  P.~Nath and R.~M.~Syed,
  Nucl.\ Phys.\ B {\bf 618}, 138 (2001).


\bibitem{Nath:2003rc}
  P.~Nath and R.~M.~Syed,
  Nucl.\ Phys.\ B {\bf 676}, 64 (2004).


\bibitem{Nath:2005bx}
  P.~Nath and R.~M.~Syed,
  JHEP {\bf 0602}, 022 (2006).





\bibitem{Aulakh:2002zr}
  C.~S.~Aulakh and A.~Girdhar,
  Int.\ J.\ Mod.\ Phys.\ A {\bf 20}, 865 (2005);
  T.~Fukuyama,
  Int.\ J.\ Mod.\ Phys.\ A {\bf 28}, 1330008 (2013);
  Z.~Y.~Chen, D.~X.~Zhang and X.~Z.~Bai,
  arXiv:1707.00580 [hep-ph].


\bibitem{Babu:2006nf}
  K.~S.~Babu, I.~Gogoladze and Z.~Tavartkiladze,
  Phys.\ Lett.\ B {\bf 650}, 49 (2007).


\bibitem{Babu:2011tw}
  K.~S.~Babu, I.~Gogoladze, P.~Nath and R.~M.~Syed,
  Phys.\ Rev.\ D {\bf 85}, 075002 (2012).


\bibitem{Enomoto:2011py}
  S.~Enomoto and N.~Maekawa,
  Phys.\ Rev.\ D {\bf 84}, 096007 (2011);
  K.~S.~Babu and R.~N.~Mohapatra,
  Phys.\ Rev.\ Lett.\  {\bf 109}, 091803 (2012).


\bibitem{Nath:2015kaa}
  P.~Nath and R.~M.~Syed,
  Phys.\ Rev.\ D {\bf 93}, no. 5, 055005 (2016).


\bibitem{Aulakh:2017aeo}
  C.~S.~Aulakh, R.~L.~Awasthi and S.~Krishna,
  arXiv:1704.04424 [hep-ph].


\bibitem{Babu:2015psa}
  K.~S.~Babu, B.~Bajc and V.~Susic,
  JHEP {\bf 1505}, 108 (2015).


\bibitem{Bajc:2013qra}
  B.~Bajc and V.~Susic,
  JHEP {\bf 1402}, 058 (2014).


\bibitem{Callaghan:2013kaa}
  J.~C.~Callaghan, S.~F.~King and G.~K.~Leontaris,
  JHEP {\bf 1312}, 037 (2013).
%


\bibitem{nudata}
  D.~N.~Spergel {\it et al.} [WMAP Collaboration],
  Astrophys.\ J.\ Suppl.\  {\bf 170}, 377 (2007);
  S.~Hannestad,
  JCAP {\bf 0305}, 004 (2003);
  Eur.\ Phys.\ J.\ C {\bf 33}, S800 (2004);
  S.~Hannestad and G.~Raffelt,
  JCAP {\bf 0404}, 008 (2004).



\bibitem{Georgi:1979df}
  H.~Georgi and C.~Jarlskog,
  Phys.\ Lett.\  {\bf 86B}, 297 (1979).



\bibitem{Froggatt:1978nt}
  C.~D.~Froggatt and H.~B.~Nielsen,
  Nucl.\ Phys.\ B {\bf 147}, 277 (1979).
  

\bibitem{harvey}
J. Harvey, P. Ramond and D. Reiss, Phys. Lett. {\bf B92},
(1980) 309; P. Ramond, R.G. Roberts,
G.G. Ross, Nucl. Phys. {\bf B406}(1993) 19; L. Ibanez and
G. G. Ross, Phys. Lett. {\bf 332}(1994) 100.


\bibitem{Ananthanarayan:1991xp}
  B.~Ananthanarayan, G.~Lazarides and Q.~Shafi,
  Phys.\ Rev.\  D {\bf 44}, 1613 (1991).
  
  
  


\bibitem{Nath:1996ft}
  P.~Nath,
  Phys.\ Lett.\ B {\bf 381}, 147 (1996);
  Phys.\ Rev.\ Lett.\  {\bf 76}, 2218 (1996).


\bibitem{Dutta:2009ij}
  B.~Dutta, Y.~Mimura and R.~N.~Mohapatra,
  Phys.\ Rev.\ D {\bf 80}, 095021 (2009);
  A.~S.~Joshipura and K.~M.~Patel,
  Phys.\ Rev.\ D {\bf 83}, 095002 (2011);
  A.~Dueck and W.~Rodejohann,
  JHEP {\bf 1309}, 024 (2013).


\bibitem{Dutta:2009bj}
  B.~Dutta, Y.~Mimura and R.~N.~Mohapatra,
  JHEP {\bf 1005}, 034 (2010).


\bibitem{Bjorkeroth:2017ybg}
  F.~Bjšrkeroth, F.~J.~de Anda, S.~F.~King and E.~Perdomo,
  arXiv:1705.01555 [hep-ph].

\bibitem{Deppisch:2016jzl}
  F.~F.~Deppisch, C.~Hati, S.~Patra, U.~Sarkar and J.~W.~F.~Valle,
  Phys.\ Lett.\ B {\bf 762}, 432 (2016);
  T.~Bandyopadhyay and A.~Raychaudhuri,
  Phys.\ Lett.\ B {\bf 771}, 206 (2017);
  Y.~Ema, K.~Hamaguchi, T.~Moroi and K.~Nakayama,
  JHEP {\bf 1701}, 096 (2017);
  B.~Sahoo, M.~K.~Parida and M.~Chakraborty,
  arXiv:1707.01286 [hep-ph];
  C.~Arbel‡ez, M.~Hirsch and D.~Restrepo,
  Phys.\ Rev.\ D {\bf 95}, no. 9, 095034 (2017);
  I.~Dorsner, S.~Fajfer and N.~Kosnik,
  Eur.\ Phys.\ J.\ C {\bf 77}, no. 6, 417 (2017);
  K.~S.~Babu, B.~Bajc and S.~Saad,
  JHEP {\bf 1702}, 136 (2017);
  M.~K.~Parida, B.~P.~Nayak, R.~Satpathy and R.~L.~Awasthi,
  JHEP {\bf 1704}, 075 (2017).


\bibitem{Weinberg:1982id}
  S.~Weinberg,
  Phys.\ Rev.\ Lett.\  {\bf 48}, 1776 (1982).
 


\bibitem{Coleman:1980aw}
  S.~R.~Coleman and F.~De Luccia,
  Phys.\ Rev.\ D {\bf 21}, 3305 (1980).
 

%

\bibitem{Cremmer:1978iv}
  E.~Cremmer, B.~Julia, J.~Scherk, P.~van Nieuwenhuizen, S.~Ferrara and L.~Girardello,
  Phys.\ Lett.\  {\bf 79B}, 231 (1978).
 


\bibitem{Cremmer:1982en}
  E.~Cremmer, S.~Ferrara, L.~Girardello and A.~Van Proeyen,
  Nucl.\ Phys.\ B {\bf 212}, 413 (1983).
  


\bibitem{Nath:1983fp}
  P.~Nath, R.~L.~Arnowitt and A.~H.~Chamseddine,
  ``Applied N=1 Supergravity,''
  ICTP Ser.\ Theor.\ Phys.\  {\bf 1} (1984).
 


\bibitem{Barbieri:1982eh}
  R.~Barbieri, S.~Ferrara and C.~A.~Savoy,
  Phys.\ Lett.\  {\bf 119B}, 343 (1982).
 


\bibitem{Ibanez:1982ee}
  L.~E.~Ibanez,
  Phys.\ Lett.\  {\bf 118B}, 73 (1982).



\bibitem{Witten:1995ex}
  E.~Witten,
  Nucl.\ Phys.\ B {\bf 443}, 85 (1995).


\bibitem{Cremmer:1978km}
  E.~Cremmer, B.~Julia and J.~Scherk,
  Phys.\ Lett.\  {\bf 76B}, 409 (1978).
 


\bibitem{Kachru:2003aw}
  S.~Kachru, R.~Kallosh, A.~D.~Linde and S.~P.~Trivedi,
  Phys.\ Rev.\ D {\bf 68}, 046005 (2003).


\bibitem{Balasubramanian:2005zx}
  V.~Balasubramanian, P.~Berglund, J.~P.~Conlon and F.~Quevedo,
  JHEP {\bf 0503}, 007 (2005).


\bibitem{AlvarezGaume:1983gj}
  L.~Alvarez-Gaume, J.~Polchinski and M.~B.~Wise,
  Nucl.\ Phys.\ B {\bf 221}, 495 (1983).
 


\bibitem{Nath:1983iz}
  P.~Nath, R.~L.~Arnowitt and A.~H.~Chamseddine,
  HUTP-83/A077, NUB-2588a.


\bibitem{Ellis:1985jn}
  J.~R.~Ellis, K.~Enqvist, D.~V.~Nanopoulos and K.~Tamvakis,
  Phys.\ Lett.\  {\bf 155B}, 381 (1985);
M.~Drees,
  Phys.\ Lett.\  B {\bf 158}, 409 (1985);
  G.~Anderson, C.H.~Chen, J.F.~Gunion, J.D.~Lykken, T.~Moroi and Y.~Yamada,
  [hep-ph/9609457].


\bibitem{nonuni2}
 A.~Corsetti and P.~Nath,
  Phys.\ Rev.\  D {\bf 64}, 125010 (2001);
U.~Chattopadhyay and P.~Nath,
  Phys.\ Rev.\  D {\bf 65}, 075009 (2002);
  A.~Birkedal-Hansen and B.~D.~Nelson,
  Phys.\ Rev.\  D {\bf 67}, 095006 (2003);
  U.~Chattopadhyay and D.~P.~Roy,
  Phys.\ Rev.\  D {\bf 68}, 033010 (2003);
  D.~G.~Cerdeno and C.~Munoz,
  JHEP {\bf 0410}, 015 (2004);
  G.~Belanger, F.~Boudjema, A.~Cottrant, A.~Pukhov and A.~Semenov,
  Nucl.\ Phys.\  B {\bf 706}, 411 (2005);
  H.~Baer, A.~Mustafayev, E.~K.~Park, S.~Profumo and X.~Tata,
  JHEP {\bf 0604}, 041 (2006);
  K.~Choi and H.~P.~Nilles
  JHEP {\bf 0704} (2007) 006;
  I.~Gogoladze, R.~Khalid, N.~Okada and Q.~Shafi,
  arXiv:0811.1187 [hep-ph];
   I.~Gogoladze, F.~Nasir, Q.~Shafi and C.~S.~Un,
  Phys.\ Rev.\ D {\bf 90}, no. 3, 035008 (2014);
    S.~Bhattacharya, A.~Datta and B.~Mukhopadhyaya,
  Phys.\ Rev.\  D {\bf 78}, 115018 (2008);
  S.~P.~Martin,
  Phys.\ Rev.\ D {\bf 79}, 095019 (2009);
   B.~Altunkaynak, P.~Grajek, M.~Holmes, G.~Kane and B.~D.~Nelson,
 JHEP {\bf 0904}, 114 (2009);
  U.~Chattopadhyay, D.~Das and D.~P.~Roy,
 Phys.\ Rev.\ D {\bf 79}, 095013 (2009);
    S.~Bhattacharya and J.~Chakrabortty,
Phys.\ Rev.\ D {\bf 81}, 015007 (2010);




\bibitem{Martin:2009ad}
  S.~P.~Martin,
  Phys.\ Rev.\ D {\bf 79}, 095019 (2009).


\bibitem{Feldman:2009zc}
  D.~Feldman, Z.~Liu and P.~Nath,
  Phys.\ Rev.\ D {\bf 80}, 015007 (2009).


\bibitem{Nilles:1982ik}
  H.~P.~Nilles,
  Phys.\ Lett.\  {\bf 115B}, 193 (1982).
 


\bibitem{Arbey:2012dq}
  A.~Arbey, M.~Battaglia, A.~Djouadi and F.~Mahmoudi,
  JHEP {\bf 1209}, 107 (2012).


\bibitem{Feldman:2006wd} 
  D.~Feldman, B.~Kors and P.~Nath,
  Phys.\ Rev.\ D {\bf 75}, 023503 (2007).



\bibitem{Pagels:1981ke}
  H.~Pagels and J.~R.~Primack,
  Phys.\ Rev.\ Lett.\  {\bf 48}, 223 (1982).


\bibitem{Weinberg:1982zq}
  S.~Weinberg,
  Phys.\ Rev.\ Lett.\  {\bf 48}, 1303 (1982).


\bibitem{Arnowitt:1992aq}
  R.~L.~Arnowitt and P.~Nath,
  Phys.\ Rev.\ Lett.\  {\bf 69}, 725 (1992).


\bibitem{Goldberg:1983nd}
  H.~Goldberg,
  Phys.\ Rev.\ Lett.\  {\bf 50}, 1419 (1983)
  Erratum: [Phys.\ Rev.\ Lett.\  {\bf 103}, 099905 (2009)].


\bibitem{Larson:2010gs}
  D.~Larson {\it et al.},
  Astrophys.\ J.\ Suppl.\  {\bf 192}, 16 (2011).
  


\bibitem{Ade:2015xua}
  P.~A.~R.~Ade {\it et al.} [Planck Collaboration],
  Astron.\ Astrophys.\  {\bf 594}, A13 (2016).
  


\bibitem{Griest:1990kh}
  K.~Griest and D.~Seckel,
  Phys.\ Rev.\ D {\bf 43}, 3191 (1991).


\bibitem{Nath:1992ty}
  P.~Nath and R.~L.~Arnowitt,
  Phys.\ Rev.\ Lett.\  {\bf 70}, 3696 (1993).
 


\bibitem{Bell:2013wua}
  N.~F.~Bell, Y.~Cai and A.~D.~Medina,
  Phys.\ Rev.\ D {\bf 89}, no. 11, 115001 (2014).


\bibitem{Baker:2015qna}
  M.~J.~Baker {\it et al.},
  JHEP {\bf 1512}, 120 (2015).


\bibitem{Weinberg:2013aya}
  D.~H.~Weinberg, J.~S.~Bullock, F.~Governato, R.~Kuzio de Naray and A.~H.~G.~Peter,
  Proc.\ Nat.\ Acad.\ Sci.\  {\bf 112}, 12249 (2014).


\bibitem{Governato}
 F.~Governato {\it et al.},
  Mon.\ Not.\ Roy.\ Astron.\ Soc.\  {\bf 422}, 1231 (2012)



\bibitem{Kim:2015yna}
  J.~E.~Kim and D.~J.~E.~Marsh,
  Phys.\ Rev.\ D {\bf 93}, no. 2, 025027 (2016);

\bibitem{Hui:2016ltb}
  L.~Hui, J.~P.~Ostriker, S.~Tremaine and E.~Witten,
  Phys.\ Rev.\ D {\bf 95}, no. 4, 043541 (2017);

\bibitem{Halverson:2017deq}
  J.~Halverson, C.~Long and P.~Nath,
  arXiv:1703.07779 [hep-ph] (to appear in PRD).


\bibitem{gross}
D. J. Gross, D.J. Harvey, E. Martinec, and R. Rohm, Phys. Rev. Lett.
{\bf 54} (1985) 502; Nucl. Phys. {\bf B256}(1985) 253; ibid. {\bf B267}
(1986) 75.



\bibitem{Horava:1995qa}
  P.~Horava and E.~Witten,
  Nucl.\ Phys.\ B {\bf 460}, 506 (1996).


\bibitem{Vafa:1996xn}
  C.~Vafa,
  Nucl.\ Phys.\ B {\bf 469}, 403 (1996).


\bibitem{Duff:2002rw}
  M.~J.~Duff,
  hep-th/0201062.


\bibitem{Acharya:2001gy}
  B.~S.~Acharya and E.~Witten,
  hep-th/0109152.


\bibitem{Acharya:2015oea}
  B.~S.~Acharya, K.~Bo?ek, M.~Crispim Rom‹o, S.~F.~King and C.~Pongkitivanichkul,
  Phys.\ Rev.\ D {\bf 92}, no. 5, 055011 (2015).


\bibitem{Douglas:2006es}
  M.~R.~Douglas and S.~Kachru,
  Rev.\ Mod.\ Phys.\  {\bf 79}, 733 (2007).



\bibitem{Candelas:1985en}
  P.~Candelas, G.~T.~Horowitz, A.~Strominger and E.~Witten,
  Nucl.\ Phys.\ B {\bf 258}, 46 (1985).

\bibitem{early-string-pheno}
A. Antoniadis, J.Ellis, J. Hagelin, and D.V. Nanopoulos, Phys. Lett.
{\bf B194}(1987)231; L.E.Ibanez, H. P. Nilles and F. Quevedo, Nucl. Phys.{\bf B307}(1988)109;
B. R. Green, K. H. Kirklin, P.J. Miron G.G. Ross,
Nucl. Phys.{\bf B292} {1987}606; R. Arnowitt and P. Nath, Phys. Rev.
{\bf D40} (1989)191;  A. Farragi, Phys. Lett. {\bf B278} (1992)131.


\bibitem{lewellen}
D.C. Lewellen, Nucl. Phys. {\bf B337}(1990)61; J. A. Schwarz, Phys. Rev.
Phys. Rev. {\bf D42}(1990)1777; S. Chaudhuri,
S.-W. Chung, G. Hockney, and J.D. Lykken, Nucl. Phys.{\bf 452}(1995)89;
G.B. Cleaver,Nucl. Phys. {\bf B456}(1995)219.


\bibitem{Kakushadze:1997mc}
  Z.~Kakushadze, G.~Shiu, S.~H.~H.~Tye and Y.~Vtorov-Karevsky,
  Int.\ J.\ Mod.\ Phys.\ A {\bf 13}, 2551 (1998).


%

\bibitem{Angelantonj:2002ct}
  C.~Angelantonj and A.~Sagnotti,
  Phys.\ Rept.\  {\bf 371}, 1 (2002).


\bibitem{Ott:2003yv}
  T.~Ott,
  Fortsch.\ Phys.\  {\bf 52}, 28 (2004).


\bibitem{Blumenhagen:2005mu}
  R.~Blumenhagen, M.~Cvetic, P.~Langacker and G.~Shiu,
  Ann.\ Rev.\ Nucl.\ Part.\ Sci.\  {\bf 55}, 71 (2005).


\bibitem{Blumenhagen:2006ci}
  R.~Blumenhagen, B.~Kors, D.~Lust and S.~Stieberger,
  Phys.\ Rept.\  {\bf 445}, 1 (2007).

\bibitem{Kors:2003wf}
  B.~Kors and P.~Nath,
  Nucl.\ Phys.\ B {\bf 681}, 77 (2004);
  P.~G.~Camara, L.~E.~Ibanez and A.~M.~Uranga,
  Nucl.\ Phys.\ B {\bf 689}, 195 (2004);
  M.~Grana, T.~W.~Grimm, H.~Jockers and J.~Louis,
  Nucl.\ Phys.\ B {\bf 690}, 21 (2004);
  D.~Lust, P.~Mayr, R.~Richter and S.~Stieberger,
  Nucl.\ Phys.\ B {\bf 696}, 205 (2004);
  D.~Lust, S.~Reffert and S.~Stieberger,
  Nucl.\ Phys.\ B {\bf 706}, 3 (2005).


\bibitem{Raby:2011jt}
  S.~Raby,
  Rept.\ Prog.\ Phys.\  {\bf 74}, 036901 (2011).
  



\bibitem{Schellekens:1989qb}
  A.~N.~Schellekens,
  Phys.\ Lett.\ B {\bf 237}, 363 (1990).


\bibitem{Dienes:1996du}
  K.~R.~Dienes,
  Phys.\ Rept.\  {\bf 287}, 447 (1997).




\bibitem{Lykken:1996fj}
  J.~D.~Lykken,
  Phys.\ Rev.\ D {\bf 54}, R3693 (1996).
 


\bibitem{Arkani-Hamed:1998rs}
  N.~Arkani-Hamed, S.~Dimopoulos and G.~R.~Dvali,
  Phys.\ Lett.\ B {\bf 429}, 263 (1998).
  

\bibitem{Antoniadis:1998ig}
  I.~Antoniadis, N.~Arkani-Hamed, S.~Dimopoulos and G.~R.~Dvali,
  Phys.\ Lett.\ B {\bf 436}, 257 (1998).
  


\bibitem{Beasley:2008dc}
  C.~Beasley, J.~J.~Heckman and C.~Vafa,
  JHEP {\bf 0901}, 058 (2009).

\bibitem{Donagi:2008kj}
  R.~Donagi and M.~Wijnholt,
  Adv.\ Theor.\ Math.\ Phys.\  {\bf 15}, no. 6, 1523 (2011).

\bibitem{Dolan:2011aq}
  M.~J.~Dolan, J.~Marsano and S.~Schafer-Nameki,
  JHEP {\bf 1112}, 032 (2011).

\bibitem{Grassi:2014zxa}
  A.~Grassi, J.~Halverson, J.~Shaneson and W.~Taylor,
  JHEP {\bf 1501}, 086 (2015).


\bibitem{Dirac:1948um}
  P.~A.~M.~Dirac,
  Phys.\ Rev.\  {\bf 74}, 817 (1948).
  


\bibitem{Schwinger:1966nj}
  J.~S.~Schwinger,
  Phys.\ Rev.\  {\bf 144}, 1087 (1966).
  


\bibitem{Kephart:2017esj}
  T.~W.~Kephart, G.~K.~Leontaris and Q.~Shafi,
  arXiv:1707.08067 [hep-ph].


\bibitem{Milton:2006cp}
  K.~A.~Milton,
  Rept.\ Prog.\ Phys.\  {\bf 69}, 1637 (2006).



\bibitem{Nath:2006ut}
  P.~Nath and P.~Fileviez Perez,
  Phys.\ Rept.\  {\bf 441}, 191 (2007).
  


\bibitem{Raby:2008pd}
  S.~Raby {\it et al.},
  arXiv:0810.4551 [hep-ph].


\bibitem{Hewett:2012ns}
  J.~L.~Hewett {\it et al.},
  arXiv:1205.2671 [hep-ex].


\bibitem{Klebanov:2003my}
  I.~R.~Klebanov and E.~Witten,
  Nucl.\ Phys.\ B {\bf 664}, 3 (2003).
  


\bibitem{Cvetic:2006iz}
  M.~Cvetic and R.~Richter,
  Nucl.\ Phys.\ B {\bf 762}, 112 (2007).


\bibitem{TheSuper-Kamiokande:2017tit}
  K.~Abe {\it et al.} [Super-Kamiokande Collaboration],
  Phys.\ Rev.\ D {\bf 96}, no. 1, 012003 (2017).


\bibitem{Abe:2011ts}
  K.~Abe {\it et al.},
  arXiv:1109.3262 [hep-ex].


\bibitem{dim5}
S. ~Weinberg, Phys. Rev. D 26, 287 (1982); 
  N.~Sakai and T.~Yanagida,
  Nucl.\ Phys.\ B {\bf 197} (1982) 533.


\bibitem{Dimopoulos:1981dw}
  S.~Dimopoulos, S.~Raby and F.~Wilczek,
  Phys.\ Lett.\ B {\bf 112} (1982) 133

\bibitem{Ellis:1981tv}
  J.~R.~Ellis, D.~V.~Nanopoulos and S.~Rudaz,
  ``Guts 3: Susy Guts 2,''
  Nucl.\ Phys.\ B {\bf 202} (1982) 43;  

\bibitem{Nath:1985ub}
  P.~Nath, A.~H.~Chamseddine and R.~Arnowitt,
  Phys.\ Rev.\ D {\bf 32} (1985) 2348;
  Phys.\ Lett.\ B {\bf 156} (1985) 215;

\bibitem{Hisano:1992jj}
  J.~Hisano, H.~Murayama and T.~Yanagida,
  Nucl.\ Phys.\ B {\bf 402} (1993) 46;

\bibitem{Lucas:1996bc}
  V.~Lucas and S.~Raby,
  Phys.\ Rev.\ D {\bf 55} (1997) 6986;

\bibitem{Babu:1998wi}
  K.~S.~Babu, J.~C.~Pati and F.~Wilczek,
  Nucl.\ Phys.\ B {\bf 566} (2000) 33;
  Phys.\ Lett.\ B {\bf 423}, 337 (1998).


  \bibitem{Goto:1998qg}
  T.~Goto and T.~Nihei,
  Phys.\ Rev.\ D {\bf 59} (1999) 115009;

\bibitem{Bajc:2002bv} 
  B.~Bajc, P.~Fileviez Perez and G.~Senjanovic,
  Phys.\ Rev.\ D {\bf 66}, 075005 (2002).



\bibitem{Ibrahim:2007fb}
  T.~Ibrahim and P.~Nath,
  Rev.\ Mod.\ Phys.\  {\bf 80}, 577 (2008).

\bibitem{Ibrahim:2000tx} 
  T.~Ibrahim and P.~Nath,
  Phys.\ Rev.\ D {\bf 62}, 095001 (2000).

\bibitem{Nath:2007eg}
  P.~Nath and R.~M.~Syed,
  Phys.\ Rev.\ D {\bf 77}, 015015 (2008).


\bibitem{Liu:2013ula}
  M.~Liu and P.~Nath,
  Phys.\ Rev.\ D {\bf 87}, no. 9, 095012 (2013).


\bibitem{Degrassi:2012ry}
  G.~Degrassi, S.~Di Vita, J.~Elias-Miro, J.~R.~Espinosa, G.~F.~Giudice, G.~Isidori and A.~Strumia,
  JHEP {\bf 1208}, 098 (2012).


\bibitem{Bednyakov:2015sca}
  A.~V.~Bednyakov, B.~A.~Kniehl, A.~F.~Pikelner and O.~L.~Veretin,
  Phys.\ Rev.\ Lett.\  {\bf 115}, no. 20, 201802 (2015).


\bibitem{Chatrchyan:2012ufa}
  S.~Chatrchyan {\it et al.} [CMS Collaboration],
  Phys.\ Lett.\ B {\bf 716}, 30 (2012).


\bibitem{Aad:2012tfa}
  G.~Aad {\it et al.} [ATLAS Collaboration],
  Phys.\ Lett.\ B {\bf 716}, 1 (2012).


\bibitem{Akula:2011aa}
  S.~Akula, B.~Altunkaynak, D.~Feldman, P.~Nath and G.~Peim,
  Phys.\ Rev.\ D {\bf 85}, 075001 (2012).



\bibitem{susy-higgs}
  H.~Baer, V.~Barger and A.~Mustafayev,
  Phys.\ Rev.\ D {\bf 85}, 075010 (2012);
   A.~Arbey, M.~Battaglia, A.~Djouadi, F.~Mahmoudi and J.~Quevillon,
  Phys.\ Lett.\ B {\bf 708}, 162 (2012);
   P.~Draper, P.~Meade, M.~Reece and D.~Shih,
  Phys.\ Rev.\ D {\bf 85}, 095007 (2012);
  M.~Carena, S.~Gori, N.~R.~Shah and C.~E.~M.~Wagner,
  JHEP {\bf 1203}, 014 (2012);
O.~Buchmueller {\it et al.},
  Eur.\ Phys.\ J.\ C {\bf 72}, 2020 (2012);
  S.~Akula, P.~Nath and G.~Peim,
  Phys.\ Lett.\ B {\bf 717}, 188 (2012);
  C.~Strege, G.~Bertone, F.~Feroz, M.~Fornasa, R.~Ruiz de Austri and R.~Trotta,
  JCAP {\bf 1304}, 013 (2013);
 K.~Kowalska, S.~Munir, L.~Roszkowski, E.~M.~Sessolo, S.~Trojanowski and Y.~L.~S.~Tsai,
  Phys.\ Rev.\ D {\bf 87}, 115010 (2013).



\bibitem{Baer:2015fsa}
  H.~Baer, V.~Barger and M.~Savoy,
  Phys.\ Scripta {\bf 90}, 068003 (2015).


\bibitem{Ibanez:2007pf}
  L.~E.~Ibanez and G.~G.~Ross,
  Comptes Rendus Physique {\bf 8}, 1013 (2007).


\bibitem{Nath:1997qm}
  P.~Nath and R.~L.~Arnowitt,
  Phys.\ Rev.\ D {\bf 56}, 2820 (1997).


\bibitem{Chan:1997bi}
  K.~L.~Chan, U.~Chattopadhyay and P.~Nath,
  Phys.\ Rev.\ D {\bf 58}, 096004 (1998);
P.~Nath and R.~L.~Arnowitt,
  Phys.\ Rev.\ D {\bf 56}, 2820 (1997);
  J.~L.~Feng, K.~T.~Matchev and T.~Moroi,
  Phys.\ Rev.\ Lett.\  {\bf 84}, 2322 (2000);
  U.~Chattopadhyay, A.~Corsetti and P.~Nath,
  Phys.\ Rev.\ D {\bf 68}, 035005 (2003);
  H.~Baer, C.~Balazs, A.~Belyaev, T.~Krupovnickas and X.~Tata,
  JHEP {\bf 0306}, 054 (2003);
  D.~Feldman, G.~Kane, E.~Kuflik and R.~Lu,
  Phys.\ Lett.\ B {\bf 704}, 56 (2011);
  S.~Akula, M.~Liu, P.~Nath and G.~Peim,
  Phys.\ Lett.\ B {\bf 709}, 192 (2012);
  G.~G.~Ross, K.~Schmidt-Hoberg and F.~Staub,
  JHEP {\bf 1703}, 021 (2017).


\bibitem{Kaufman:2015nda}
  B.~Kaufman, P.~Nath, B.~D.~Nelson and A.~B.~Spisak,
  Phys.\ Rev.\ D {\bf 92}, 095021 (2015).



\bibitem{Chatterjee:2012qt}
  R.~M.~Chatterjee, M.~Guchait and D.~Sengupta,
  Phys.\ Rev.\ D {\bf 86}, 075014 (2012).

\bibitem{Nath:2016kfp}
  P.~Nath and A.~B.~Spisak,
  Phys.\ Rev.\ D {\bf 93}, no. 9, 095023 (2016).

\bibitem{Choudhury:2017lxb} 
  A.~Choudhury, K.~Kowalska, L.~Roszkowski, E.~M.~Sessolo and A.~J.~Williams,
  Universe {\bf 3}, no. 2, 41 (2017).
  
\bibitem{Aboubrahim:2017aen}
  A.~Aboubrahim, P.~Nath and A.~B.~Spisak,
  Phys.\ Rev.\ D {\bf 95}, no. 11, 115030 (2017).

\bibitem{Francescone:2014pza} 
  D.~Francescone, S.~Akula, B.~Altunkaynak and P.~Nath,
  JHEP {\bf 1501}, 158 (2015);
  D.~Feldman, Z.~Liu and P.~Nath,
  Phys.\ Rev.\ Lett.\  {\bf 99}, 251802 (2007).
  
 \bibitem{NU}
D. Matalliotakis and H. P. Nilles, Nucl. Phys. {\bf B435}, 115(1995);
M. Olechowski and S. Pokorski, Phys. Lett. {\bf B344}, 201(1995);
N. Polonski and A. Pomerol, Phys. Rev.{\bf D51}, 6532(1995);
  P.~Nath and R.~Arnowitt,
  Phys.\ Rev.\  D {\bf 56}, 2820 (1997);
   E.~Accomando, R.~L.~Arnowitt, B.~Dutta and Y.~Santoso,
  Nucl.\ Phys.\ B {\bf 585}, 124 (2000);
J.~R.~Ellis, K.~A.~Olive and Y.~Santoso,
  Phys.\ Lett.\  B {\bf 539}, 107 (2002);
H.~Baer, A.~Mustafayev, S.~Profumo, A.~Belyaev and X.~Tata,
  JHEP {\bf 0507}, 065 (2005);
    U.~Chattopadhyay and D.~Das,
  Phys.\ Rev.\ D {\bf 79}, 035007 (2009).



\bibitem{Aboubrahim:2017wjl}
  A.~Aboubrahim and P.~Nath,
  arXiv:1708.02830 [hep-ph] (to appear in PRD).

\bibitem{Kane:2009kv} 
  G.~Kane, P.~Kumar and J.~Shao,
  Phys.\ Rev.\ D {\bf 82}, 055005 (2010)
  doi:10.1103/PhysRevD.82.055005
  [arXiv:0905.2986 [hep-ph]].

\bibitem{Ellis:2017vpy} 
  S.~Ellis,
  ``Phenomenological Aspects of Supersymmetry with Heavier Scalars,''
  (U Mich Thesis 2017).


\bibitem{Pradler:2006qh}
  J.~Pradler and F.~D.~Steffen,
  Phys.\ Rev.\ D {\bf 75}, 023509 (2007).


\bibitem{Patrignani:2016xqp}
  C.~Patrignani {\it et al.} [Particle Data Group],
  Chin.\ Phys.\ C {\bf 40}, no. 10, 100001 (2016).




\end{thebibliography}
\end{document}